\documentstyle[draft,epsf]{mn}

\def\gsim{\mathrel{\raise0.35ex\hbox{$\scriptstyle >$}\kern-0.6em 
\lower0.40ex\hbox{{$\scriptstyle \sim$}}}}
\def\lsim{\mathrel{\raise0.35ex\hbox{$\scriptstyle <$}\kern-0.6em 
\lower0.40ex\hbox{{$\scriptstyle \sim$}}}}
\def\hi{{\rm H\,{\sc i} }}

\def\rr{{\rm R$_{25}$ }}
\def\rrns{{\rm R$_{25}$}}
\def\hii{{\rm H\,{\sc ii} }}

\def\hh{{\rm H$_2$ }}

\title[The properties of spiral galaxies]{The properties 
of spiral galaxies: confronting 
hierarchical galaxy formation models with observations}

\author[Bell et al.]
{
Eric F. Bell$^1$, Carlton M. Baugh$^2$, Shaun Cole$^2$, Carlos S.
Frenk$^2$ and \vspace{0.15cm} \\
\vspace{0.2cm} {\LARGE Cedric G. Lacey$^{2,3}$} \\
$^1$ Max Planck Institut f\"ur Astronomie, K\"onigstuhl 17, D-69117, 
	Heidelberg, Germany; 
        {\rm \texttt{bell@mpia.de}}\\
$^2$ Department of Physics, University of Durham, 
	South Road, Durham DH1 3LE, UK\\
$^3$ CRAL Observatorie de Lyon, 9 Avenue Charles Andr\'e, F 69561
	Saint Genis Laval Cedex, France
}

\begin{document}
\date{\fbox{{\sc MNRAS, accepted}: \today}}

\label{firstpage}

\maketitle

\begin{abstract}
We compare the properties of local spiral galaxies with the
predictions of the Cole et al. semi-analytic model of hierarchical
galaxy formation, in order to gain insight into the baryonic
processes, such as gas cooling and star formation, that were
responsible for shaping these galaxies.  On the whole, the models
reproduce the properties of present-day spirals rather well, including
the trend in scale-size with luminosity, the width of the scale-size
distribution, the tight gas fraction--surface brightness and gas
fraction--star formation history correlations, the
metallicity--magnitude correlation, and the present-day star formation
rates and stellar mass-to-light ratios.  Of special note is our
demonstration that, once the effects of dust and variations in stellar
populations have been taken into account, published spiral galaxy
scale-size distributions derived from optical data (with widths
$\sigma \sim 0.3$) can be reconciled with the width ($\sigma \sim
0.5$) of the stellar mass scale-size distribution predicted by the
semi-analytic model. 

There are some illuminating discrepancies between
the observations and the Cole et al. model predictions.  The
model colours of luminous spiral galaxies are somewhat too blue and
those of faint galaxies somewhat too red, most likely indicating
shortcomings in the way that gas is accreted by spiral galaxies.  
Furthermore, the model produces too few luminous
spiral galaxies. These
difficulties could be resolved by altering the way in which gas cooling is
treated or, perhaps, by adopting a higher baryon fraction 
and invoking galactic 
``superwinds.''  Secondly, stellar mass-to-light ratios are found to 
be as high as observations permit. Yet, typically 60 per cent of the
mass in the inner half-light radius of the model galaxies is dark.
This results in an offset between the model and observed spiral galaxy
luminosity-linewidth relation. This could be resolved by substantially
reducing the mass of baryons which make it into a galaxy disc (with an
attendant decrease in stellar mass-to-light ratio), or by modifying
the assumed dark matter profile to include less dark matter in the
inner parts.
\end{abstract}


\section{Introduction}

Understanding the place of spiral galaxies in hierarchical
models of galaxy formation is of special importance for two 
main reasons.
Firstly, one of the principal, justifiable, assumptions underpinning
many studies of 
galaxy formation in a Cold Dark Matter (CDM) context
is that all galaxies form initially as discs 
(e.g., Baugh, Cole \& Frenk 1996; Abadi et al.\ 2003; 
although see Nulsen \& Fabian 1997 for a different
view).  Only subsequently are the signatures of discs
erased by ubiquitous process of violent galaxy mergers \cite{barnes92}.
Thus, understanding galaxy discs is a prerequisite for understanding
all galaxies.
Secondly, because of their ongoing star formation and disc morphology
(indicating a lack of {\it recent} violent mergers), spiral disc galaxies
give a unique insight into the formation and evolution  
of ``quiescent'' galaxies.  
This quiescent evolution depends strongly on 
certain processes and the 
way in which they are modelled, such as gas cooling, 
star formation (SF) and feedback.  
These phenomena, because of their complex nature, are amongst the most 
difficult to incorporate in the models. 
In this work, we hope to improve our understanding of these processes by 
comparing the properties of local spiral galaxies with model spiral galaxies 
from the semi-analytic Cole et al.\ (2000; CLBF hereafter) simulations.

Semi-analytic models attempt to capture the essential features 
of galaxy formation in a CDM cosmology 
using simple but effective parameterisations of the relevant physical
processes. 
In this way, it becomes possible to establish the relative importance
of various  processes in determining galaxy properties, and 
to identify specific areas of uncertainty 
\cite{white78,cole91,white91,lacey91,kauf93,cole94,somer99,c00,vdb02,benson03}.
The results of state-of-the-art, but computationally expensive, 
$N$-body and gas dynamical simulations are used to motivate the 
prescriptions employed in semi-analytic models to describe processes such 
as the the growth of dark matter haloes and the radiative cooling of gas 
\cite{cole96,benson01a,springel01,murali02,helly02a,helly02b}.  

The global properties of galaxies are generally well reproduced by
semi-analytic models.  Here, we highlight just a few of the many
successes of this approach.  Galaxy sizes, morphological mixes, 
luminosity functions, the slope
and scatter of the Tully-Fisher (TF) relation, and the star formation
history (SFH) of the Universe are all reproduced well
\cite[CLBF]{white91,kauf93,cole94,baugh96,baugh98,somer99,somer01}.
For spheroids, the black hole mass--luminosity/velocity dispersion
relation and the colour--magnitude relation can be interpreted
within this framework \cite{kc98,kauf00}.  Furthermore,
when combined with high resolution $N$-body simulations, these models
provide a good description of the clustering properties of galaxies at
both low and high redshift
\cite{governato98,baugh99,kauffmann99,benson00,benson01b,norberg01,mathis02}.

However some predictions of the semi-analytic models are not in such
good agreement with the observations.  No model has yet been able to match
the luminosity function and the TF relation zero-point {\it
simultaneously} if the disc
circular velocity is constructed in a way that accounts for 
the gravitational effect of the baryon collapse
(Kauffmann et al.\ 1993; Cole et al.\ 1994; Heyl et
al.\ 1995; Somerville \& Primack 1999; 
van Kampen, Jimenez \& Peacock 1999; CLBF).
This problem has also been seen in $N$-body
gas-dynamic simulations (Navarro \& Steinmetz 2000; Eke, Navarro \& Steinmetz
2001; N.\ Katz, 2002, private communication) and `analytical' 
studies \cite{mo00,vdb00,yang03}.   
In the model of CLBF, who give most weight to
matching the luminosity function when setting model parameter values,
this problem manifests itself as a velocity offset of $\sim$30 per
cent in the zero point of the TF relation (when the disc circular
velocity at the half mass radius is used: when the halo circular
velocity is used the offset disappears).  Furthermore, it is unclear
whether it is possible to reproduce the colour-magnitude relation of
early-type galaxies without drastically upsetting the agreement with
the observed optical and near-IR luminosity functions
\cite[CLBF]{kc98}.  The colour-magnitude relation of spiral galaxies
is also poorly-reproduced \cite{vdb02}; this is a correlation that we
investigate in depth in this paper.\footnote{Other approaches
which include some of the CDM-motivated scaling laws but
none of the hierarchical merging characteristic of the 
semi-analytic approach
may not suffer from this limitation
\protect\cite{mmw,avila,boissier}.  Because the modelling
philosophy differs from that presented here, we do
not discuss these `hybrid' approaches further, 
referring the interested reader to the above works directly.}
  In addition, the radial profiles
of model spiral galaxies are roughly exponential, but inevitably have
large, prominent central concentrations which are not seen in many
later-type spiral galaxies \cite{dalc97,vdb01}.  \footnote{There are
other challenges facing those wishing to understand galaxy 
formation.  Most notable among these is the `angular momentum
catastrophe', the tendency of gas-dynamical simulations of
galaxy formation to produce galaxies with up to an order of 
magnitude too little angular momentum to match the present-day 
galaxy population \cite{white93,stein99}.  This problem 
does not directly affect the semi-analytical approach because angular momentum
is assumed to be conserved during galaxy formation, although
the validity of this assumption is certainly questioned by these
simulation results.}

In this paper, we concentrate on the global properties of local spiral galaxy discs
in the fiducial model of CLBF.  By focusing on local discs, we can develop an 
understanding of how best to parameterise complex baryonic processes
such as star formation and gas cooling, not only for this model, 
but for all models of this type.
The plan of the paper is as follows.  In section \ref{sec:model},
we briefly discuss the model.  
In section \ref{sec:obs}, we outline which model observables
we test against data, and describe the observational 
datasets used in these comparisons.
In section \ref{sec:phys}, we compare model 
predictions for physical properties of spiral galaxies (such as luminosity, size
or gas fraction) with the observations.  In section \ref{sec:colour},
we compare the optical--near-IR colours of a diverse sample of 
spiral galaxies with the models, to elucidate their 
metallicities and SFHs.  In section \ref{sec:sfl}, we
examine the observed and model present day star formation rates (SFRs).  
We discuss maximum-disc constraints on the 
model stellar mass to light ratios (M/Ls) in section \ref{sec:ml}. 
In section \ref{sec:spin}, we revisit the 
scale-size distribution of spirals (which was already discussed
in CLBF), in the light of new data. 
In section \ref{sec:disc}, we present a 
discussion of the major discrepancies between the models and 
the observations, identifying a few possibilities for reconciling
the two.  Finally, 
we summarise our conclusions in section \ref{sec:conc}.
Novel features of this paper include: a discussion of  
comparisons between the observed spiral galaxy scale-size distributions 
in different passbands with the spin parameter distribution predicted by 
galaxy formation models; an explicit comparison 
between the model and observed SF laws; 
a presentation of a novel method for comparing models with inhomogeneously-selected 
datasets; a discussion of age and metallicity effects on the spiral galaxy colour-magnitude
relations; and, a demonstration of the utility of comparing spiral
galaxy M/Ls and their luminosity-linewidth relation to models, simultaneously.
We set $H_0 = 100h\,{\rm km\,s^{-1}\,Mpc^{-1}}$ hereafter unless stated
otherwise.

\section{The galaxy formation model} \label{sec:model}

The semi-analytic model used in this paper is comprehensively 
described by CLBF. 
A cursory outline of the model is as follows: 
(i) The growth of dark matter haloes through mergers and accretion 
is followed using a Monte-Carlo algorithm. 
(ii) The lowest mass branches of the dark matter halo merger tree 
are filled with hot gas, which is assumed to be shock-heated to the 
virial temperature of the halo.
(iii) Gas is allowed to cool by atomic radiative processes. 
(iv) Cooled gas is turned into stars according to an adopted star formation 
law. 
(v) The return of energy into the cooled gas, by stellar winds or supernovae, 
reheats a fraction of this gas, thereby providing a feedback mechanism to 
regulate star formation; this is particularly effective in low circular 
velocity haloes. 
(vi) Galaxies merge with the central galaxy in a halo if their dynamical 
friction timescale is shorter than the lifetime of the host dark 
matter halo. A violent merger is marked by a burst of star formation and the 
destruction of any stellar disc.
The semi-analytic model of CLBF follows the chemical evolution of gas and 
stars and uses the Bruzual \& Charlot stellar population synthesis model 
to compute the luminosity of a galaxy in any desired passband. 

Here, we provide only a brief description of the 
physical processes which we show have a direct bearing on the 
properties of disc galaxies: gas cooling
and disc formation, star formation and feedback, and 
the chemical and stellar evolution of galaxies.
We discuss in section \ref{sec:disc} what changes are produced 
in the model predictions if
some of the prescriptions adopted by CLBF are varied.
We note that the fiducial model of CLBF assumes a baryon 
density a factor of two lower than subsequent measurements have 
indicated \cite{omeara01,spergel03}.  We discuss 
the effects of cosmic baryon density later in section \ref{sec:disc}.

\subsection{Gas cooling and disc formation} \label{sec:gas}

Motivated by the results of extensive $N$-body simulations, the
dark matter is assumed to follow a 
Navarro, Frenk \& White \shortcite[NFW hereafter]{nfw95,nfw} density profile.
Haloes also have a hot gas component which is assumed to have
been shock-heated to the virial temperature of that halo during that
halo's collapse. Motivated by the simulations 
of Eke et al. \shortcite{eke98}, we assume that the hot gas follows a 
$\beta$-model profile, $\rho_{\rm gas} \propto 1/(r^2 + r_{\rm core}^2)$,
where $r_{\rm core}$ is the core radius of the gas distribution.
The core radius starts as a constant fraction of the halo size,
and grows as low entropy gas cools (the growth is roughly a factor
of four by the present day).  The rate at which 
gas cools is density and metallicity-dependent, and
is quite sensitive to this hot gas density profile: this 
means that different assumptions about the density profile can 
produce significant 
changes in the formation history and angular momentum content
of present-day discs (although comparison of gas-dynamical models
and the semi-analytic prescription have given a reassuring overall agreement; 
Benson et al.\ 2001; Helly et al.\ 2002b).  

The dark matter and hot gas have angular momentum, imparted
by tidal torques.  The magnitude of the dark matter and 
gas angular momentum are 
characterised by the dimensionless spin parameter $\lambda = 
J|E|^{1/2}/(GM^{5/2})$, where $M$, $J$, and $E$ are the
total halo mass, angular momentum and energy.  Motivated
by the $N$-body simulations of Cole \& Lacey \shortcite{cole96}, 
we adopt a log-normal distribution of $\lambda$ with mean
0.039 and logarithmic width 0.53.  This distribution 
agrees with more recent determinations \cite{lemson99,bullock01}.
In order to 
estimate the radial distribution of angular momentum in 
the halo and gas, we choose to adopt a constant mean 
rotation velocity within the halo as a function of
radius, broadly consistent with the results of
Cole \& Lacey \shortcite{cole96}.
During cooling and collapse we assume that the 
gas conserves angular momentum and forms a disc. 
The assumption of conservation of angular momentum leads to good
agreement between predicted and observed galaxy sizes 
(e.g., Mo, Mao \& White 1998, CLBF).  
However, this assumption is not supported by 
gas-dynamical simulations of galaxy formation which 
include radiative cooling, but no effective stellar feedback. 
In these gas-dynamical simulations, the gas transfers most of its angular 
momentum to the dark haloes, resulting in galaxy sizes that are 
much smaller than observed \cite{white93,stein99}.  
Feedback from star formation may promote conservation of angular momentum, leading
to better agreement with observed disc sizes (e.g., Weil, Eke
\& Efstathiou 1998; Thacker \& Couchman 2001). 
In order to be consistent with the observed distribution 
of disc sizes, we assume that angular momentum is conserved; 
we leave the mechanism by which it is conserved 
as an as yet unanswered question.

In this paper, we compare the properties of observed field
spiral galaxies with {\it central} halo disc-dominated galaxies 
in haloes less massive than $10^{14} h^{-1}$ M$_{\sun}$.
This selection is made because observationally, our sample galaxies are typically
reasonably isolated, or are at least the dominant members of
their group, and $10^{14} h^{-1}$ M$_{\sun}$ is the halo mass of 
a large group or small cluster.
We do not compare the observations with model satellite galaxies.
Gas is assumed to cool only onto the galaxy at the centre of 
a halo (see van Kampen, Jimenez \& Peacock 1999 
for an alternative cooling model).  
Thus, satellite galaxies are starved of a fresh infalling gas supply, and
quickly fade and redden.  It is impossible to test the accuracy of this 
assumption using observations of what are more than likely to 
correspond to central halo galaxies in our model. 
Including satellite galaxies in the comparisons with the observed data 
leaves our conclusions regarding luminous discs unchanged; however, 
it would introduce a substantial population of non-star-forming, 
red discs at lower luminosities (in clear conflict with the data).

\subsection{Star formation and feedback} \label{subsec:sf}

Gas is converted into stars at the rate 
$\psi = M_{\rm cold}/\tau_*$, where $\psi$ is
the SFR, $M_{\rm cold}$ is the mass 
of cold gas, and $\tau_*$ is the timescale for
star formation.  This timescale is related to the
disc dynamical timescale, $\tau_{\rm disc}$, and the disc potential:
$\tau_* = \epsilon_*^{-1} \tau_{\rm disc} (V_{\rm disc}/200\,{\rm km\,s}^{-1})^{\alpha_*}$, 
where $\epsilon_*$ is a star formation efficiency,
$V_{\rm disc}$ is the circular velocity of the disc at the half-mass 
radius, and ${\alpha_*}$ governs the galaxy mass dependence in the SF law.  
If $\alpha_* = 0$, the star formation timescale is simply proportional
to the disc dynamical
timescale.  CLBF's fiducial model adopts ${\alpha_*} = -1.5$ to 
suppress star formation in smaller haloes, ensuring a good match to 
the observed high gas fractions of less luminous galaxies.

In this model, energy from supernovae and stellar winds ejects
cold gas and metals from the disc into the halo.  
The mass of cold gas ejected into the halo is given by 
$\dot{M}_{\rm eject} = \beta \psi$, where
$\beta = (V_{\rm disc}/V_{\rm hot})^{-\alpha_{\rm hot}}$; 
$V_{\rm hot} = 200\,{\rm km\,s^{-1}}$ 
and $\alpha_{\rm hot} = 2.0$ are the parameters that govern the strength 
of the feedback process.
The cold gas takes its share of metals with it into 
the halo.  None of the freshly-synthesised metals are 
ejected directly into the halo.  This freshly-ejected 
gas plays no role in gas cooling until is incorporated into a
hot gas halo as a result of a galaxy merger.   
These parameter values are quoted for the fiducial model; 
in particular, $\alpha_{\rm hot} = 2.0$ was adopted so that the 
model predictions are in agreement with both the slope of the faint 
end of the luminosity function and the shape of the TF relation.  
Adopting a larger value of $\alpha_{\rm hot}$ leads to excessive 
curvature in the TF relation (Cole et al.\ 1994).
If feedback was not included, the luminosity
function would be much steeper than that estimated from observations 
\cite{norberg02,vdb03,benson03}.

\subsection{Chemical evolution, stellar populations and dust}
 \label{subsec:sp}

Chemical evolution of the gas and stars is followed assuming that
the initial reservoir of gas is free of metals.  Metals
are formed by stars and incorporated into the cold gas using
the instantaneous recycling approximation \cite{tinsley80}.  Furthermore,
some of the cold gas may be ejected into the halo, taking with it its share
of metals.  

The luminosities, spectra and integrated colours of the stellar
components are modelled using 
the stellar population synthesis (SPS) models of Bruzual \& Charlot
(in preparation), which are described in more detail in 
Lui, Charlot \& Graham \shortcite{liu00} and 
Charlot \& Longhetti \shortcite{charlot01}.  The models
allow for a wide range of stellar ages and metallicities. 
Briefly, stellar populations with older ages or higher metallicities
are redder and fainter (because of lower mean stellar temperatures)
than populations which are younger, or have lower metallicity.
Uncertainties in the predictions of the colours, luminosities 
and spectra of SPS models for older stellar populations are
$\sim$10 per cent \cite{charlot96}. 

We assume a universal stellar initial mass function (IMF): the one 
adopted by Kennicutt \shortcite{k83}.
This IMF is similar to a Salpeter \shortcite{sp}
IMF at the high mass end (and therefore has similar colours), 
but has fewer low mass stars (and therefore has a lower stellar 
M/L, by around a factor of two).  In order to match
the zero point of the luminosity function, the stellar M/L
is increased by a factor of 1.38 (which was
ascribed to a contribution to the total stellar mass 
from brown dwarves, but in reality
simply reflects uncertainty in the shape of the IMF at low masses).
This IMF, with this modification of stellar M/L, is consistent
with estimates of and limits on stellar M/Ls (Bell \& de Jong 2001; 
CLBF; see also section \ref{sec:ml}).  

Dust affects the luminosities and colours of stellar populations.  
Dust is taken into account using the model of Ferrara 
et al.\ \shortcite{ferrara99}, which assumes a smooth dust distribution 
mixed in with the stellar disc.
The overall optical depth of the galaxy is computed self consistently 
using the total mass of metals per unit area in the cold gas.  
The luminosities and colours of the model galaxies include  
the effects of dust, assuming a random distribution
of galaxy inclinations $i \le 60$ degrees (in the case of 
the predictions presented in this paper).  We choose
this inclination distribution to mimic as closely as possible
the observational selection of the 
Bell \& de Jong \shortcite{papii} sample.  The 
optical--near-infrared luminosity and colour
effects of dust for $i \le 60$ degrees galaxies are relatively minimal in this 
model, and our results are not significantly affected by assuming
a clumpy dust distribution \cite{granato00}, 
or by neglecting the effects of dust entirely 
($\la 0.1$ mag and $\la 0.3$ mag respectively, 
for the most dusty, usually most luminous model galaxies; these
effects are smaller than the dust vectors in Figs.\ 
\ref{fig:colcolmag}--\ref{fig:colcolfg}).

\section{Comparing the model with observations} \label{sec:obs}

We select disc-dominated galaxies from the fiducial model of CLBF
with $K$-band bulge-to-disc ratios $B/D \le 1$ to compare with
observations (see Baugh et al. 1996 for a plot of $B/D$ versus T-type).  
Different aspects of this comparison allow us to focus in detail 
on the following model features.
\begin{itemize}
\item Differences in the luminosity function probe the physics of 
  galaxy mergers and gas heating and cooling processes. 
\item The colours of spiral galaxies are driven by their
  stellar populations, which in turn allow us to test the gas cooling, 
  star formation and feedback prescriptions.
\item The SFRs of spiral galaxies primarily reflect the adopted star formation
  prescription.
\item The observational limits on the M/Ls of spiral galaxies 
  depend principally on the IMF.  When combined with 
  the TF relation, further insight can be gleaned about the IMF and the fraction 
  of dark matter contained within the inner parts of galaxies.
\item The sizes of spiral galaxies primarily 
  constrain the adopted distribution of angular momentum or, equivalently, the 
  halo spin parameter.
\end{itemize}

In this paper, we compare the model predictions with a number of datasets.
\begin{enumerate}
\item An inhomogeneously-selected sample of 121 galaxies from 
Bell \& de Jong \shortcite{papii}.  This sample lacks a well-defined
selection criterion; however, it has high-quality optical--near-IR
colours at the disc half-light radius, accurate physical parameters,
and is very diverse, allowing the exploration of the properties of a
huge range of galaxy discs.  It is a combination of three samples: de
Jong \shortcite{djiv}, which is diameter-limited; Tully et al.\
\shortcite{tv}, which is volume-limited (with magnitude and surface
brightness cutoffs); and Bell et al.\ \shortcite{papi}, which is
selected to have low surface brightness (LSB), and as wide a range of
luminosity, size and colour as possible.  Due to the inhomogeneous
selection, the selection of model galaxies for comparison with these
observations is made by matching with the observed $K$-band  
surface brightness and $K$-band luminosity (and,
therefore, to within a factor of two, the stellar surface density and
mass; Bell \& de Jong 2001) of each galaxy in the
sample.  This is discussed in more detail later.
\item An inhomogeneously-selected sample of 
star-forming spiral galaxies from Kennicutt \shortcite{k98}.  These 
galaxies have accurate H$\alpha$ surface photometry and gas 
surface density maps, allowing  the comparison of model
and observed SF laws.  Unfortunately, this dataset lacks 
surface brightness measurements, meaning that the selection 
technique described in (i) above cannot 
be attempted.  A much cruder surface brightness 
limit is applied to the model galaxies in this case.
\item We compare model stellar M/Ls with data from a complete sample
of Ursa Major Cluster galaxies \cite{verheijen}.
\end{enumerate}
In addition, we discuss observational data from some other papers
\cite{dejong00,cross02,kauf02}, and other model/observational
comparison papers  
\cite[CLBF]{dejong00,vdb02}, when we feel that their discussion will
prove useful. 

\section{Matching model galaxies to an inhomogeneous observational
database: the physical properties of the Bell \& de Jong sample}
 \label{sec:phys}

\begin{figure*}
\begin{minipage}{175mm}
\begin{center}
  \leavevmode
  \epsffile{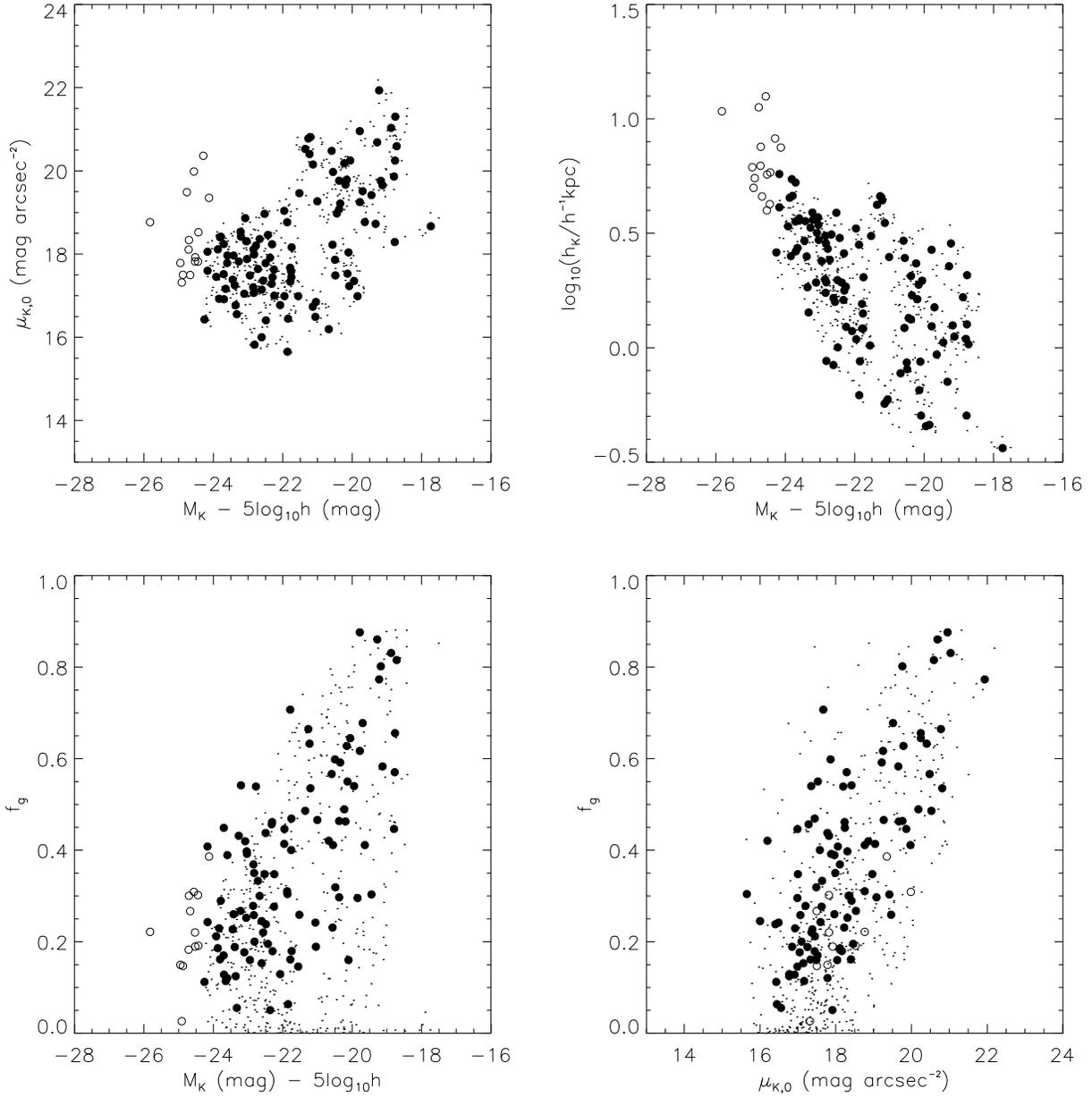}
\end{center}
\caption{A comparison of the disc parameters from the model
galaxies (dots) and the combined dataset of 
Bell \& de Jong \protect\shortcite{papii}.
Model galaxies are selected for comparison in this Figure, and
in Figs.\ \protect\ref{fig:colcolmag}--\protect\ref{fig:colcolfg}, 
by their proximity to observed galaxies in the $K$-band central surface
brightness--absolute magnitude plane (upper left panel).
Observed galaxies are denoted by solid circles if five or more 
model galaxies fall within a locus of radius 0.4 mag 
in the $\mu_{K,0}$--($M_K - 5 \log_{10} h$) plane.
For these galaxies, we show 5 randomly selected model galaxies within
this locus.  Observed galaxies are denoted by open circles
if fewer than five galaxies satisfy this criterion; in this case, 
no model galaxies are shown.  Also shown are
comparisons between the distributions of $K$-band disc scale lengths and
magnitudes (upper right panel), estimated total gas fraction and $K$-band
absolute magnitude (lower left panel), and total gas fraction and $K$-band
central surface brightness (lower right panel).
}
\label{fig:phys}
\end{minipage}
\end{figure*}

\begin{figure}
\begin{center}
  \leavevmode
  \epsfxsize 8cm
  \epsfysize 8cm
  \epsffile{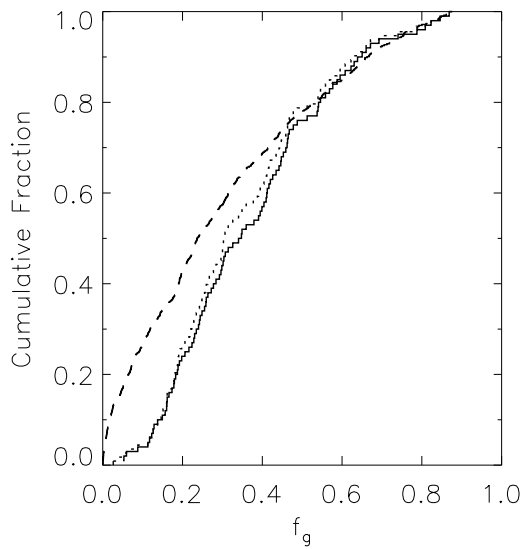}
\end{center}
\caption{ We show the cumulative distribution of the 
gas fractions of all observed galaxies (dotted line), only those 
observed galaxies with matching model galaxies (solid line),
and the model galaxies constrained to have 
similar surface brightnesses and magnitudes (dashed line).
}
\label{fig:gascum}
\end{figure}

In this section, we introduce the method used to select 
model galaxies to compare with the inhomogeneously-selected
dataset of Bell \& de Jong \shortcite{papii}.  We also discuss 
the luminosities, surface brightnesses and gas fractions of the
model galaxies and examine how well they match the observations.
We adopt total $K$-band magnitudes corrected for 
Galactic foreground extinction following Schlegel, Finkbeiner \&
Davis \shortcite{sfd}.  Central surface brightnesses 
$\mu_{K,0}$ are extrapolated disc-only central surface brightnesses,
corrected to face-on assuming the optically-thin limit \cite{papii}.
We do not account for dust extinction owing to dust in 
the individual galaxies, as this is accounted for in the galaxy 
modelling.

In Fig.\ \ref{fig:phys}, we show the physical properties (the $K$-band
absolute magnitude $M_K$, the $K$-band central surface brightness
$\mu_{K,0}$, the $K$-band disc scale length $h_K$, and the estimated
total gas fraction, including Helium and molecular hydrogen, $f_g$) of
the combined sample of 121 galaxies from Bell \& de Jong
\shortcite{papii}.  We have chosen to take a simple approach to the
selection of model galaxies for comparison with the data.  For each
observed galaxy, we searched for model galaxies within a locus of 0.4
mag radius in the $M_K - \mu_{\rm K,0}$ plane.  (The results are
insensitive to the exact choice of size of the locus.)  If five or
more model galaxies are found within this locus, then we select five
of them to compare with the data.  We show observed galaxies matched
in this way using solid circles in Fig.\ \ref{fig:phys} (and Figs.\
\ref{fig:colcolmag}--\ref{fig:colcolfg} and \ref{fig:bd100}).  Some
galaxies do not have five model analogues; these galaxies are left
unmatched, and are denoted by open circles.  Using this scheme, we
ensure that we compare observed galaxies with model galaxies with
appropriate luminosities and disc central surface brightnesses.

Examining the upper left-hand panel of Fig.\ \ref{fig:phys}, 
we see that there are 15
luminous, large scale-length, high surface brightness 
(HSB; with $\mu_{\rm K,0} \sim 18$ mag\,arcsec$^{-2}$,
corresponding roughly to $\mu_{\rm B,0} \sim 21.5$ mag\,arcsec$^{-2}$)
galaxies which remain unmatched in the $M_K - \mu_{\rm K,0}$ plane.
There are very few model disc galaxies in this area of the plane; 
de Jong \& Lacey \shortcite{dejong00} also noted a lack of  
very luminous spiral galaxies (see their Fig.\ 10).
This is not a trivial discrepancy to rectify;  
this point is discussed further in section \ref{sec:disc}.
This is also reflected in the scale-size distribution plane (upper right
panel of  Fig.\ \ref{fig:phys}).  

The gas fractions of the observed and model galaxies are compared
in the lower panels of Fig.\ \ref{fig:phys}.  
In Fig.\ 9 of CLBF, a reasonable match between the models
and data as a function of $B$-band absolute magnitude was demonstrated,
although with an excess of model galaxies with low gas fractions
(however, gas-poor satellite galaxies were included in this comparison). 
Yet, $K$-band luminosities more directly
reflect stellar mass \cite{ml}, making a comparison 
in terms of $K$-band instructive.  We find a good match
between the the {\it trends} in gas fraction with 
$K$-band absolute magnitude and disc central surface brightness.
There is a hint of an excess of low gas-fraction galaxies (as seen in CLBF).  
In Fig.\ \ref{fig:gascum}, we show the cumulative
distribution of observed galaxy gas fractions (solid line) and
the model galaxies (dashed line).  The dotted line shows
the cumulative distribution of gas fractions for all observed
galaxies, including those without matching model analogues.
There is a clear but subtle
excess of model galaxies at low gas fractions. A Kolmogorov-Smirnov 
test indicates that the distributions are different at much 
greater than 99.9 per cent confidence.  Overall, however, it 
is fair to say that the gas fractions of the model galaxies
are in reasonable agreement with the observations, with only
a small excess of low gas fraction model galaxies.

Of particular interest is
the tightness in the gas fraction--surface brightness correlation.
The models do a reasonable job of reproducing the 
gas-fraction--magnitude and magnitude--surface brightness correlations
(the left-hand panels of Fig.\ \ref{fig:phys}).  However, when a 
scattered gas-fraction--magnitude correlation is convolved with 
a highly-scattered magnitude--surface brightness correlation, it 
is not obvious that a tight relation between surface brightness
and gas fraction should result.  Yet, the model
gas fraction--surface brightness correlation is tighter than
the gas fraction--magnitude correlation, in agreement
with the data.

\section{Spiral galaxy colours and SFHs} \label{sec:colour}

\begin{figure*}
\begin{minipage}{175mm}
\begin{center}
  \leavevmode
  \epsffile{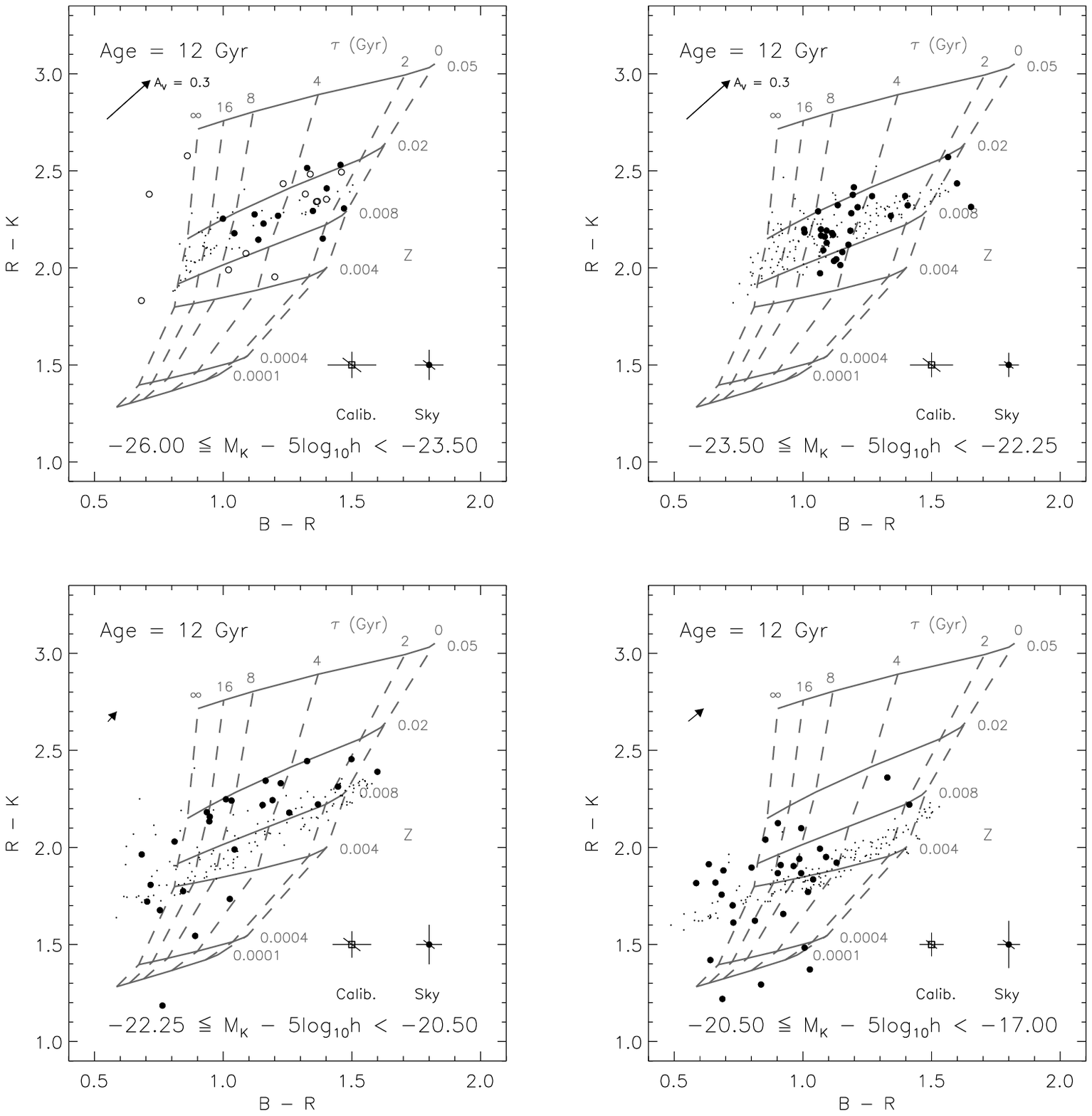}
\end{center}
\caption{A comparison of the optical--near-IR colours of 
model galaxies (points) with the colours at the half-light radius of
galaxies (solid circles for matched galaxies, open circles for
unmatched galaxies) from Bell \& de Jong \protect\shortcite{papii},
binned by $K$-band absolute magnitude.  Overplotted are stellar
population models from Bruzual \& Charlot (in preparation).  We also
show calibration and sky subtraction uncertainties; the diagonal lines
represent the effect of a $1 \sigma$ shift in the $R$-band photometry.
Overplotted also are the effects of foreground screen (upper panels)
and Triplex (lower panels) dust models.  In the Triplex model, the
dust vector denotes reddening at the half light radius for a central
$\tau_V = 2$ galaxy.  The lower left (right) panel shows a Triplex
model with a Milky Way (Small Magellanic Cloud Bar) extinction curve.
}
\label{fig:colcolmag}
\end{minipage}
\end{figure*}

\begin{figure*}
\begin{minipage}{175mm}
\begin{center}
  \leavevmode
  \epsffile{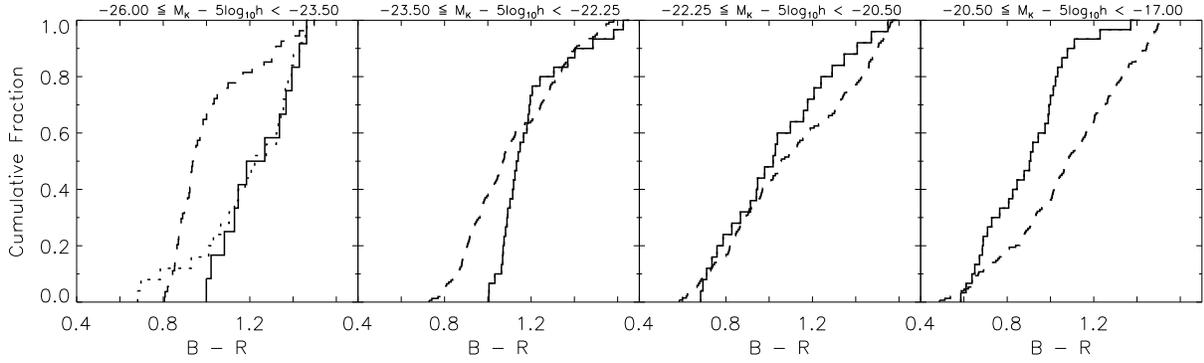}
\end{center}
\caption{Cumulative distribution of the $B-R$ colours
of observed and model galaxies, in four different magnitude bins (where
galaxy luminosity decreases to the right).  
The cumulative distributions of $B-R$ colours
of observed
galaxies with hierarchical galaxy matches 
are shown by solid lines, and the corresponding
distribution for all galaxies is shown by a dotted line.
In all but the left-most panel all galaxies have model
galaxy matches, therefore the dotted and solid lines overlap.
The model galaxy distribution is shown as a dashed line.
}
\label{fig:colcum}
\end{minipage}
\end{figure*}

\begin{figure*}
\begin{minipage}{175mm}
\begin{center}
  \leavevmode
  \epsffile{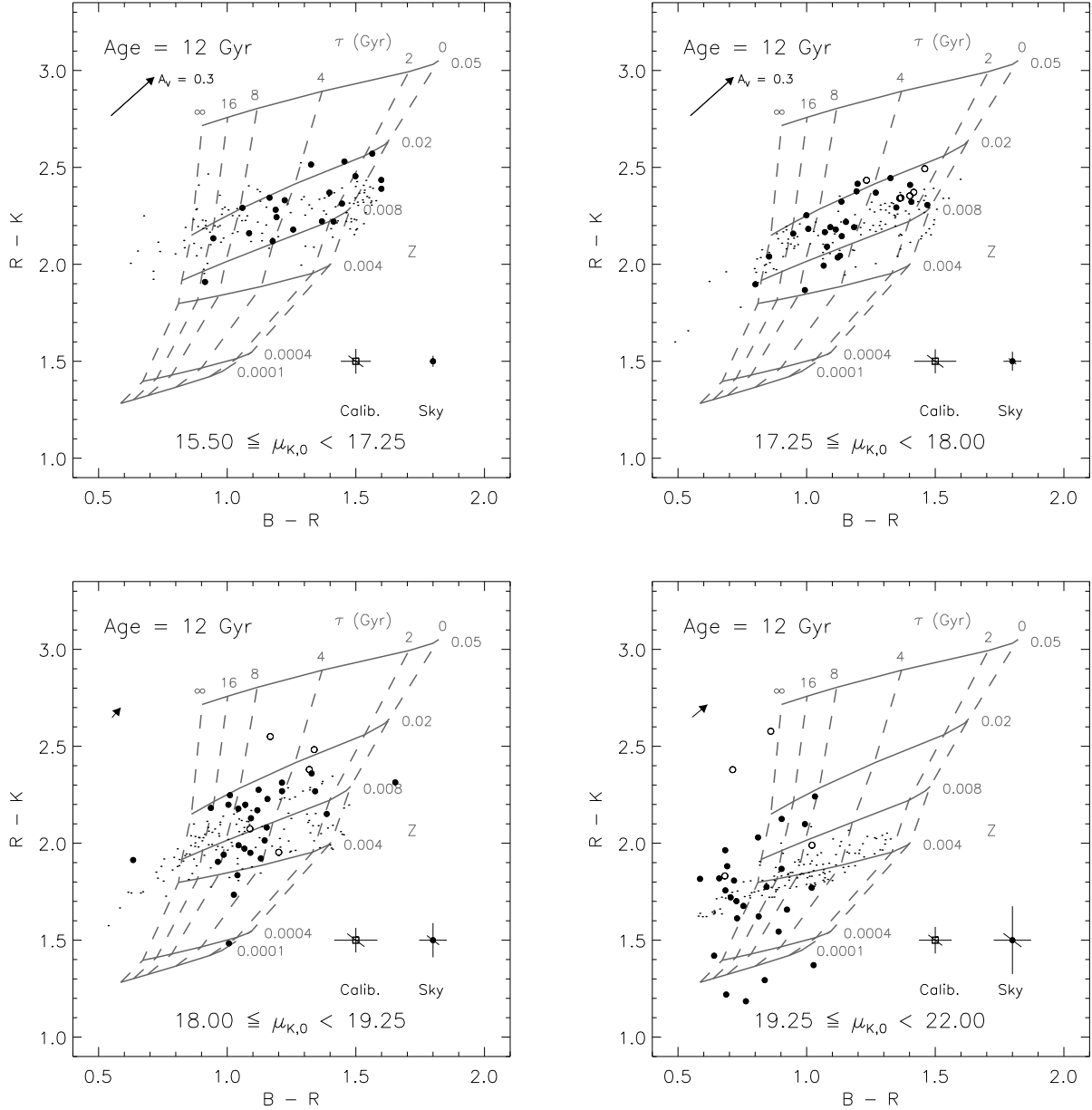}
\end{center}
\caption{Same as Fig.\ \protect\ref{fig:colcolmag}, but binned by 
$K$-band central surface brightness.
}
\label{fig:colcolcsb}
\end{minipage}
\end{figure*}

\begin{figure*}
\begin{minipage}{175mm}
\begin{center}
  \leavevmode
  \epsffile{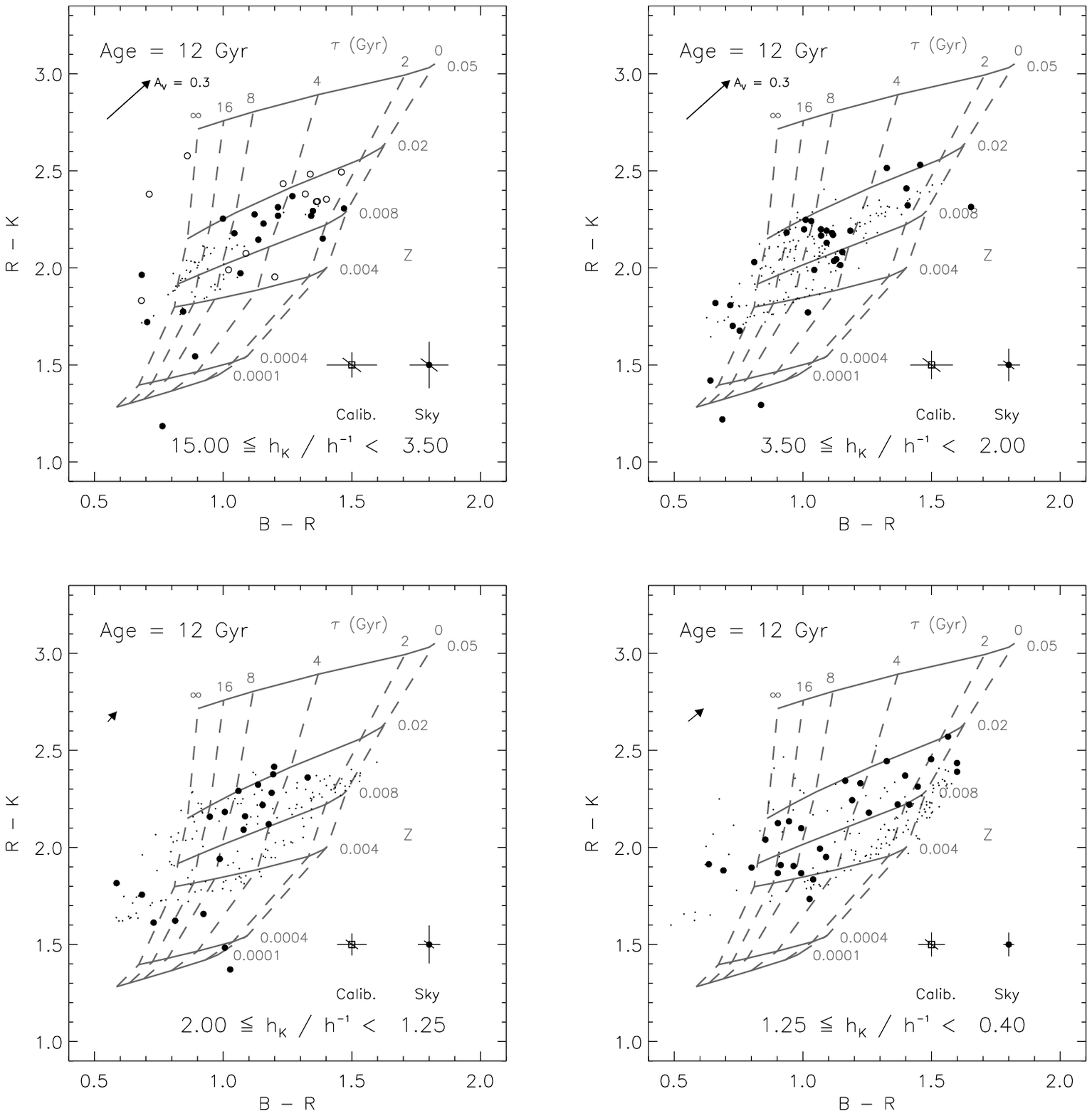}
\end{center}
\caption{Same as Fig.\ \protect\ref{fig:colcolmag}, but binned by 
$K$-band disc scale length (in $h$\,kpc).  There are fewer than 
expected galaxies in the upper left panel because of 
a slight bias towards lower scale lengths in the model galaxy matches
to the observed galaxies and inhomogeneities in the distribution 
of observed galaxy scale lengths (see the upper right-hand
panel of Fig.\ \ref{fig:phys}). 
}
\label{fig:colcolscal}
\end{minipage}
\end{figure*}

\begin{figure*}
\begin{minipage}{175mm}
\begin{center}
  \leavevmode
  \epsffile{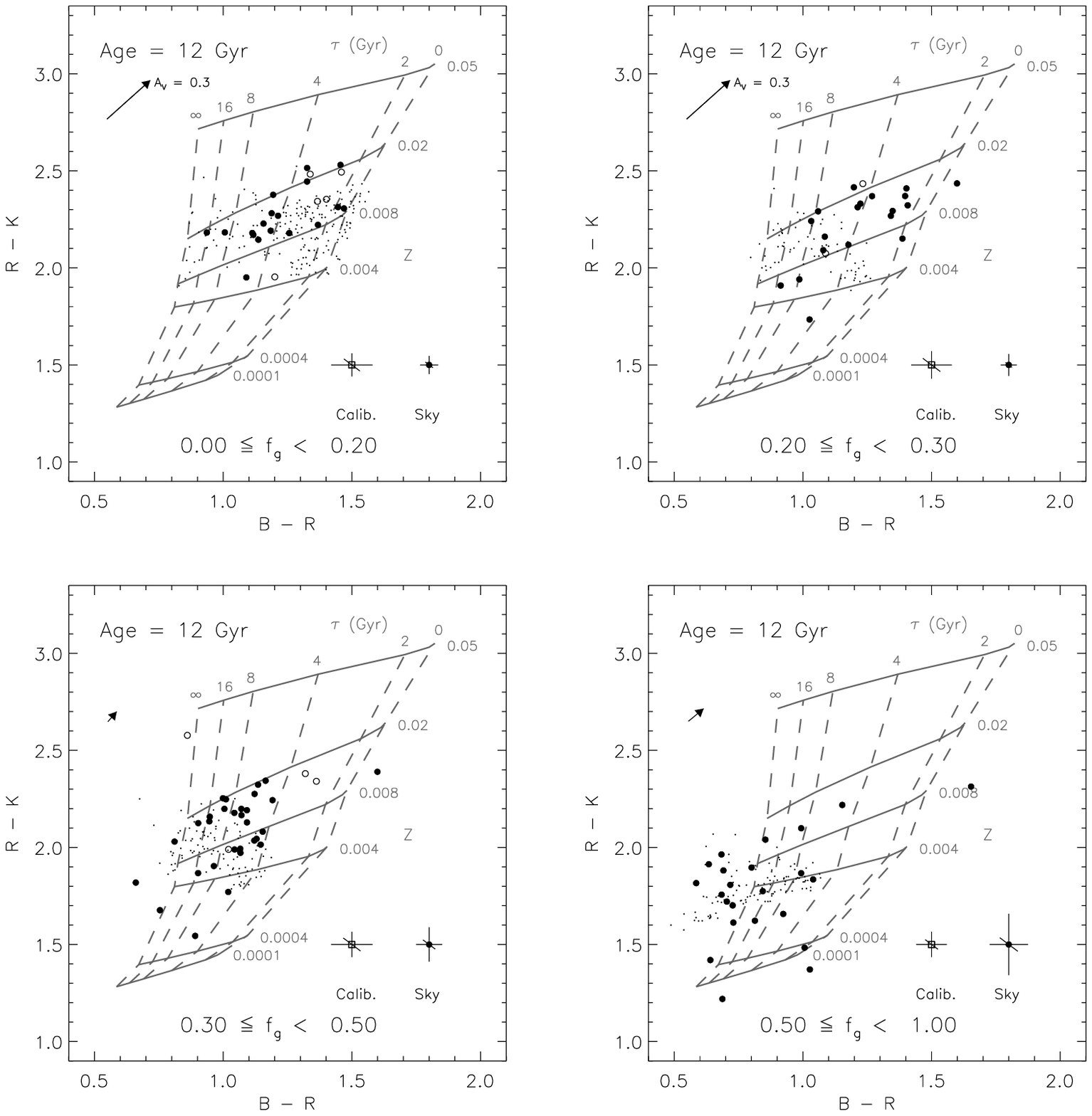}
\end{center}
\caption{Same as Fig.\ \protect\ref{fig:colcolmag}, but binned by 
estimated total gas fraction.  Because of the slight
model galaxy offset to lower gas fractions, the low gas fraction
bin is over-populated by model galaxies, compared to the observed
sample.  }
\label{fig:colcolfg}
\end{minipage}
\end{figure*}

The optical--near-IR colours of stellar populations with ongoing star
formation (such as those in spiral galaxies) can potentially provide a 
powerful insight into the SFHs and metallicities of those populations.
Essentially, optical colours are sensitive to the position
of the main sequence turn-off, whereas near-IR colours are more
sensitive to the properties of the red giants and asymptotic giant
branch stars.  
Using the optical and near-IR colours together, it is possible to 
differentiate between the effects of metallicity and, to first order, 
the ratio of young ($<$2 Gyr) to old stars (which can be taken as an age, 
or a kind of birthrate parameter; de Jong 1996; Bell \& de Jong 2000).  
The same principle underlies the estimation of ages and metallicities 
for elliptical galaxies, 
this time involving age- and metallicity-sensitive spectral
lines \cite{harald00,trager00}.

We compare the stellar populations of observed and model 
spiral galaxies using optical--near-IR colour-colour plots 
(Figs.\ \ref{fig:colcolmag}--\ref{fig:colcolfg}).  
We show the $B - R$ against $R - K$ colours measured at the half-light 
radius for galaxies from the sample of Bell \& de Jong \shortcite{papii}
(111 out of the 121 galaxies have $B$, $R$ and $K$-band data).
Radially-resolved colours were not used as CLBF treat their model
galaxies as consisting of a bulge and disc component only.  
We compare with observed colours at the half-light radius
because they are more accurate than extrapolated, total colours.
We construct half-light radius colours for the model galaxies by 
adding the correct amount of ``contamination'' from
the bulge at the disc half-light radius to the disc colours (a 
$\la 0.1$ mag effect).
The colours and luminosities include the effect of smoothly distributed
dust, for a spiral galaxy with a randomly-distributed $i \le 60$ degrees, 
following the prescription of CLBF 
(see their section \ref{subsec:sp}).  
Solid circles denote galaxies with at least five model galaxy
matches in the $M_K - \mu_{K,0}$ plane; open circles
denote unmatched galaxies (see section \ref{sec:phys}).

To give an impression of the trends expected in optical--near-IR 
colours resulting from changes in galaxy age and metallicity, we 
overplot stellar population models from Bruzual \& Charlot (in preparation),
as described in more detail by Liu, Charlot \& Graham (2000).  
The model of Kodama \& Arimoto \shortcite{ka97}, or 
the multi-metallicity {\sc P\'egase} stellar population model
(see Fioc \& Rocca-Volmerange 1997 for an earlier version of this
model) give
similar optical--near-IR colours.  
The stellar population models, like the models of CLBF,
adopt the IMF used by Kennicutt \shortcite{k83}.  
These models assume an exponentially
decreasing SFR with an $e$-folding time $\tau$,
$\psi \propto e^{-t/\tau}$;  star formation is assumed to start 12 Gyr ago. 
Models with the same metallicity but 
different $\tau$ are connected by solid lines; models with 
different metallicities and the same $\tau$ are connected by dashed
lines.  The optical--near-IR colours are insensitive to modelling
assumptions, such as the choice of IMF, the time at which star 
formation starts, and to reasonable variations in SFH \cite{papii,ml}.
Also overplotted are foreground dust screen models
(upper panels), and the more realistic Triplex 
models (which have a mixed star/dust geometry;
Disney, Davies \& Phillipps 1989) with Milky Way
(lower left panel) and Small Magellanic Cloud Bar (lower
right panel) dust.  Further description of the dataset, the stellar population
models, model assumptions, and the dust models is
given in Bell \& de Jong \shortcite{papii}.

Model galaxies, selected by their proximity to observed
galaxies in the $M_K - \mu_{K,0}$ plane (again, 5 are plotted 
for every matched, observed galaxy), are denoted by dots.  
In Fig.\ \ref{fig:colcolmag}, we show the trend in optical--near-IR 
colour with $K$-band absolute magnitude, where bright galaxies are 
in the upper left-hand panel, going through to faint galaxies in 
the lower right-hand panel.  
The $R - K$ colours of the model galaxies 
match the observed colours well.  To the extent
that one believes that $R-K$ colours are a reasonable metallicity
indicator\footnote{The main difficulty with this proposition
is dust.  While the models suggest that the effects of 
dust on the optical--near-IR colours will be modest, observational confirmation 
remains elusive.  Work is ongoing with the Two Micron All 
Sky Survey and Sloan Digital Sky Survey to test the 
role of dust in the optical and near-IR in more detail.}, this indicates 
that the metallicity--luminosity relation for model spiral galaxies 
is a good match to the observed metallicity--luminosity relation.
In contrast, CLBF found that the metallicity--luminosity
relation predicted by the model was shallower than the 
gas metallicity--luminosity relation from Zaritsky, 
Kennicutt \& Huchra \shortcite{zkh94}.
There are no systematic trends in the model between the gas and stellar 
metallicities that would generate a difference in the slopes of 
the gas and stellar metallicity--magnitude relations.  
We note, however, that there is a disagreement between the 
observational datasets.
Fig.\ 16 of Bell \& de Jong \shortcite{papii}
compares metallicities derived from the optical--near-IR colours
with gas metallicities from Zaritsky et al.\ \shortcite{zkh94}, 
amongst other studies.  Where the colour-based metallicities are
reasonably reliable (at metallicities $\ga -$1 dex), 
Bell \& de Jong \shortcite{papii} find that the colours suggest
a shallower metallicity-luminosity relation and 
somewhat lower metallicities than the strong-line
derived \hii region abundances (see, e.g., 
Kennicutt, Bresolin \& Garnett 2003 for a discussion of the
limitations of strong-line \hii region abundances).
Thus, we conclude that the metallicity--luminosity
relation of the CLBF model is in agreement with the data, to 
within the systematic uncertainties.

In contrast, the trends in observed $B - R$ colour 
with absolute magnitude are relatively poorly matched by the models (recall
that $B - R$ colour is more sensitive to SFH than metallicity; 
Fig.\ \ref{fig:colcolmag}).  We explore this in more detail
in Fig.\ \ref{fig:colcum}, where we show the cumulative
distribution of observed galaxies with model matches (solid lines)
against the distribution of model galaxies (dashed lines), 
as a function of absolute magnitude.  
At high luminosities, model galaxies
are $\sim 0.3$ mag bluer in $B - R$ colour than the observed 
galaxies, indicating that the luminous model galaxies have too 
much ongoing star formation.  The distributions of model and observed
galaxy colours are inconsistent at much greater than 99.9 per cent
confidence.
At lower luminosities, model galaxies cover the entire range of 
$B - R$ colours almost uniformly.  This matches the observed
dataset at moderate luminosity (e.g., for the interval
$-22.25 \le M_K - 5 \log_{10} h < -20.5$,
where the model and observed distributions are very similar, 
differing at only 33 per cent confidence).
However, at the lowest luminosities, the observed galaxies tend to have 
$B - R \la 1$, unlike the model galaxies which extend up to 
$B - R \sim 1.5$ (indicating that many faint galaxies have 
too little ongoing star formation in the model).  The distributions
of observed and model colours differ at 99.95 per cent confidence. 
This model shortcoming could be related to a
number of issues; we discuss this in more depth in section \ref{sec:disc}.

In Fig.\ \ref{fig:colcolcsb}, we show trends in optical--near-IR
colour with $K$-band central surface brightness.  The model
optical--near-IR colours match the observed colours reasonably well.
It appears that the models produce somewhat too large
a range in $B - R$ colour at any given surface brightness: however,
a Kolmogorov-Smirnov test indicates similar distributions in model
and observed
$B-R$ colour, except for the lowest surface brightness bin, where
the models are too red at 99.9 per cent confidence
(the cumulative distributions are not shown for brevity).  The 
trend in model $R - K$ colour with surface brightness matches
the observations reasonably well.  One interesting discrepancy
is the rather wider range in $R - K$ colour shown by the observed LSB
galaxies.  The most luminous LSB galaxies are unmatched in 
the model (open circles), and we do not discuss them further here.
However, less luminous LSB galaxies are observed to 
have a `tail' down to very blue colours: $R - K \la 1.5$.  
A value of $R - K$ this low
is very difficult to produce from the stellar population synthesis
models \cite{papii}, and therefore this cannot be reproduced by the
model of CLBF.  This problem may indicate a limitation in the SPS models
at the very lowest metallicities, or could be due to the inevitable
difficulty in accurately measuring the $K$-band magnitudes of 
very blue LSB galaxies (reflected in the sky subtraction
error bars).

Observationally, there is no evidence for a trend in 
SFH with disc scale-size; see Fig.\ \ref{fig:colcolscal}.
The distributions of $B-R$ colours for small $h_K \le 3.5 h$\,kpc model 
galaxies are not significantly different from the observed
galaxies.
Nevertheless, there is a tendency for small model galaxies to
have a wide range in metallicities at a given age, and a
wide range of ages (to first order, $B - R$ colour) in 
an absolute sense.   
Larger model galaxies are substantially
less abundant (cf. the number of unmatched, large scale
length, observed galaxies), and have a narrow range of metallicities at
a given age, and a narrow range of absolute ages. 
The distribution of model $B-R$ colours of these largest galaxies 
differs from the observations
at greater than 99.9 per cent confidence (again, the cumulative
distributions are not shown for brevity).

The model colours, binned by gas fraction 
(Fig.\ \ref{fig:colcolfg}), are a good match to the observations,
with the model and observed galaxies never differing
at more than 99 per cent confidence.
The overall correlation between age and metallicity (as probed 
by the colours) and gas fraction are as expected: old, metal-rich
(and therefore highly evolved) galaxies are poor in gas, 
whereas younger, more metal-poor (therefore less evolved)
galaxies are rich in gas.   

\section{The star formation law} \label{sec:sfl}

\begin{figure*}
\begin{minipage}{175mm}
\begin{center}
  \leavevmode
  \epsffile{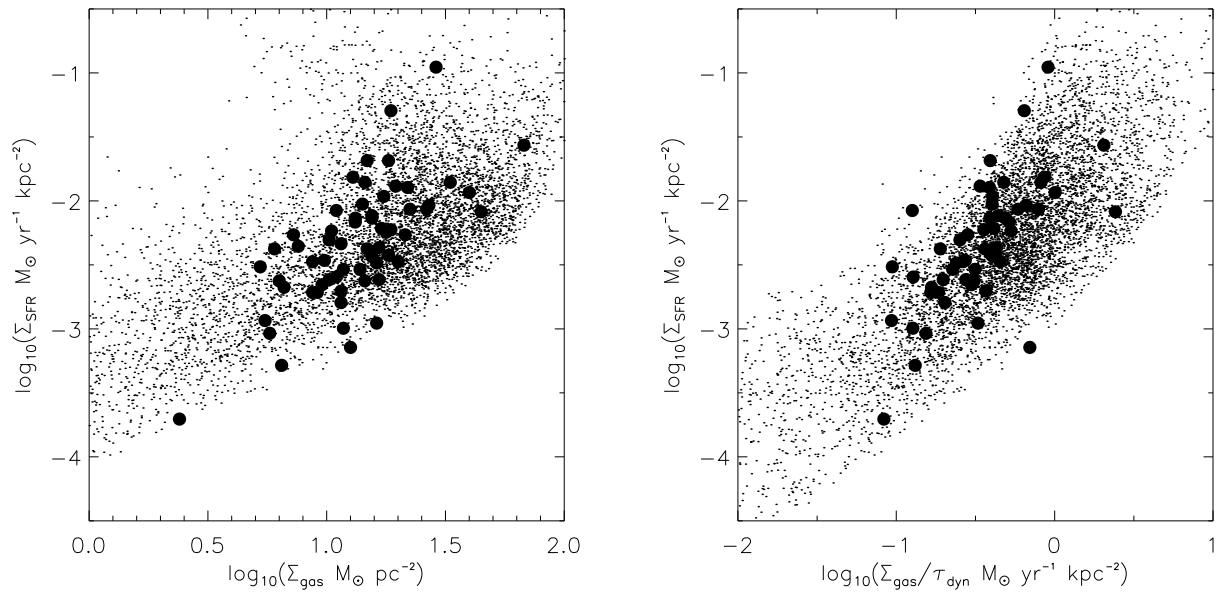}
\end{center}
\caption{ Comparison of the star formation `laws' of the 
model galaxies (dots) and observed galaxies taken from 
Kennicutt (1998; solid circles).  The left panel
shows the SFR surface density, averaged within \rrns, plotted against 
the average
gas surface density within \rrns.  The right panel shows the
SFR surface density within \rr against the 
ratio of the gas surface density within
\rr to the disc orbital timescale at \rr (as defined in the text).  
Model galaxies were selected to have $B$-band central 
surface brightnesses brighter than 22.5 mag\,arcsec$^{-2}$; 
this cutoff gives a rough approximation to the selection effects
that operate on the sample of Kennicutt \protect\shortcite{k98}.
Changing this surface brightness cutoff moves the lower 
envelope of the model galaxies towards higher
SFR densities at a given gas mass for a higher
surface brightness cutoff.
}
\label{fig:sfl}
\end{minipage}
\end{figure*}

Kennicutt \shortcite{k98} derived correlations between the 
SFR per unit area, $\Sigma_{\rm SFR}$, and {\it i)} the gas surface density,
and {\it ii)} the ratio of the gas surface density to dynamical time, 
$\Sigma_{\rm gas}/\tau_{\rm dyn}$. 
Here, we test these data against the predictions of the CLBF model, 
using Kennicutt's \shortcite{k98} observational dataset, 
converted to ${\rm H}_0 = 100$\,km\,s$^{-1}$\,Mpc$^{-1}$
in Fig.\ \ref{fig:sfl}.
Total gas surface densities within \rr were estimated from the \hi and 
\hh densities, multiplied by a factor of 1.35 to account for He.  
SFR surface densities within \rr (which include the effects of an 
average 1.1 mag. of extinction; Kennicutt 1983)
were reduced by a factor of 0.7 to convert from the 
Salpeter \shortcite{sp} IMF adopted by Kennicutt \shortcite{k98}
to the Kennicutt \shortcite{k83} IMF plus 38 per cent brown 
dwarf fraction adopted in this work.  

Model galaxy \rr values were estimated assuming exponential discs and 
$r^{1/4}$ law bulge profiles.  Dynamical times at \rr were
estimated using $\tau_{\rm dyn} = 9.78 \times 10^8 {(R_{25}/{\rm kpc})
/(V_{\rm disc}/{\rm km\,s^{-1}})}\,{\rm yr}$, where $V_{\rm disc}$ 
is the circular velocity of the disc in the disc plane
at the disc half-mass radius due to the gravity of the disc, 
bulge and dark halo components (Appendix C of CLBF).  The rotation 
curves of most galaxies are quite flat outside the half disc mass
radius, making our approximation $V_{\rm disc} \sim V_{25}$, where
$V_{25}$ is the circular velocity at \rrns, not a particularly poor one.
The time taken to complete one orbit is 2$\pi \tau_{\rm dyn}$.  
Observationally, only 60 per cent of the gas content of 
spiral galaxies lies inside $R_{25} \sim R_{\rm star-forming}$
\cite{martin01}.
Since the models do not predict where gas should be within 
the model galaxies, we adopt a simplistic reduction of the model
gas mass by 40 per cent before constructing the gas density.

It is apparent that the match between the 
model galaxy SFRs (points) 
and the observations (solid circles) is not unreasonable, when expressed
either in terms of the gas surface density or in terms of 
the ratio of the gas surface density and the disc orbital timescale.
The clumping of model points with low SFR at higher gas densities
is a possible discrepancy.  However, because it is impossible
to reproduce Kennicutt's \shortcite{k98} galaxy selection, it is difficult
to draw firm conclusions from the density of model points in certain
areas of the plot.  The lower envelope of the model points also
depends somewhat on the surface brightness cutoff
adopted when selecting the model galaxies. We adopt a 
$B$-band central surface brightnesses cutoff of 22.5 mag\,arcsec$^{-2}$.
This surface brightness limit is appropriate, as LSB
galaxies \cite{impey97}
would typically not make it into the NGC (from which most galaxies in 
the sample of Kennicutt 1998 are drawn).
Changing this surface brightness cutoff moves the lower 
envelope of the model galaxies towards higher 
SFR densities at a given gas mass for a higher 
surface brightness cutoff.

\begin{figure}
\begin{center}
  \leavevmode
  \epsfxsize 8cm
  \epsfysize 8cm
  \epsffile{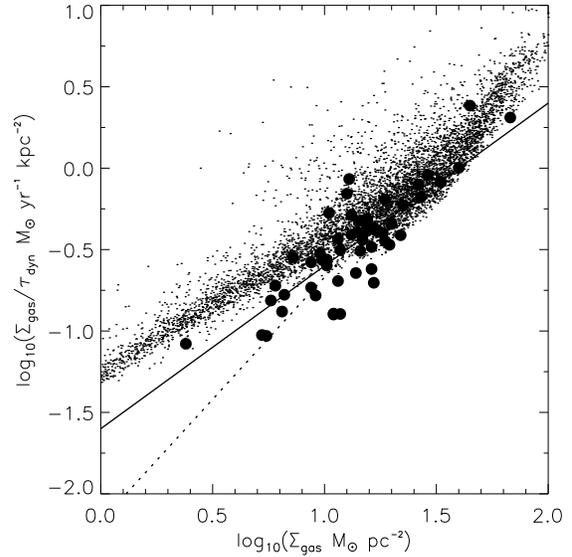}
\end{center}
\caption{ The ratio of the gas surface density within \rr
to $\tau_{\rm dyn}$ at \rrns, plotted against gas surface density.
It is apparent that the difficulty in discerning
between a `Schmidt' SF law and a dynamical time SF law is 
essentially unavoidable: both real and model galaxies
show a tight correlation between these two quantities.
The solid line denotes a fit to the observational data with a slope of unity
(RMS of 0.16 dex), and
the dotted line denotes a fit to the data with a slope of 1.5 
(RMS of 0.20 dex).
}
\label{fig:dens}
\end{figure}

The reasonable match with the right hand panel of Fig.\ \ref{fig:sfl} 
is expected: 
the SF law adopted in the fiducial model has 
the star formation timescale $\tau_* \propto \tau_{\rm dyn}\,V_{\rm disc}^{-1.5}$,
and SFR $\psi = M_{\rm gas}/\tau_*$.  In the absence of 
the velocity-dependent part of the SF law, the 
model galaxies in the right hand panel would follow a  
straight line, with some spread due to the scatter in 
the ratio of the half-mass radius to \rrns.
The distribution of model galaxies in that panel
should therefore be a straight line with a slope of unity and a zero-point
that depends relatively weakly on circular velocity.
However, the reasonable match with the left hand 
panel is less trivial: {\it a priori}, it is not clear that 
a good match with a SF law defined in terms of $\tau_{\rm dyn}$
would also result in a gas-density dependent Schmidt \shortcite{schmidt} 
SF law. 

The origin of the similarity between the right and left hand panels of
Fig.\ \ref{fig:sfl} is a close correspondence between the ratio of the
gas surface density to the dynamical time and the gas surface density
for both the real and model galaxies (Fig.\ \ref{fig:dens}).  This is
driven primarily by the gas density term in each quantity (the solid
line denotes the fit to the data with a slope of unity).  However, 
signatures of the dynamical time term are visible, especially in the 
distribution of the model points.  In particular, the dynamical
time is largest for galaxies with gas surface densities
between 10\,M$_{\sun}$\,pc$^{-1}$ and 30\,M$_{\sun}$\,pc$^{-1}$, 
and lower for galaxies with higher and lower gas densities within
\rrns.  At the low density end, galaxy discs are smaller and 
typically low surface brightness, thus \rr is relatively small.  
These galaxies are also dark matter dominated, thus driving down
the dynamical time, and driving up 
the ratio of gas surface density to dynamical time.  At the high
density end, the large surface density of baryons (and the attendant
increase in dark matter density through the adiabatic response
of the dark matter to the large baryon density) causes
these galaxies to have shorter dynamical times.  This can
be illustrated very simply by neglecting the influence of dark
matter, assuming gas fractions of order unity, and assuming that
galaxies have identical \rrns; then $\tau_{\rm dyn} \propto
\Sigma_{\rm gas}^{-1/2}$, where $\Sigma_{\rm gas}$ is the gas surface density
within \rrns.  This would give $\Sigma_{\rm gas}/\tau_{\rm dyn}
\propto \Sigma_{\rm gas}^{3/2}$ (as
denoted by the dotted line).  Whatever the driving mechanism is
that causes (weak) departures from a linear relationship between
$\Sigma_{\rm gas}$ and $\Sigma_{\rm
gas}/\tau_{\rm dyn}$, it is clear that 
SF laws formulated using either parameter
are practically indistinguishable using present-day SFRs alone.

\section{Stellar mass-to-light ratios}
	\label{sec:ml}

One of the most serious shortcomings of semi-analytic galaxy
formation models is their inability to match the luminosity function
and the zero-point of the Tully-Fisher (TF) relation simultaneously
when the rotation velocity of the disc is properly constructed
(e.g., White \& Frenk 1991, Kauffmann et al.\ 1993; Cole et al.\ 1994;
Heyl et al.\ 1995; Somerville \& Primack 1999; CLBF).  
CLBF constrain their models to reproduce the
$b_J$-band and $K$-band luminosity functions; however, these models
fail to match the zero-point of the TF relation (when disc circular
velocities are used the offset is $\sim$1 mag, or 30 per cent in terms
of circular velocity, in the sense that galaxies are either too faint
or they rotate too quickly).  A fresh insight into the problem can be
obtained by considering the mass-to-light (M/L) ratios of spiral
galaxies.  The zero-point of the TF relation is intimately related to
stellar M/Ls: for example, larger stellar M/Ls than inferred from
observation would lead to fainter galaxies, resulting in an offset
from the zero-point of the TF relation.  The stellar M/Ls of the models
were discussed briefly in section 7.7 of CLBF and were found to be
roughly consistent with observations.  In this section, we discuss the
stellar M/L of model and observed discs in greater detail; we discuss
the relationship to the TF relation in more detail in section
\ref{disc:tf}.

Observationally, stellar M/Ls are difficult to estimate reliably.  
M/L values in the solar neighbourhood are inferred 
using dynamical arguments and by direct accounting
of the stellar mass (e.g.\ Binney \& Tremaine 1987).  
Estimates of M/L in spiral galaxies can be obtained from 
(challenging) measurements of velocity dispersion,
under some assumptions about the disc vertical structure
\cite{bottema93,bottema99,swaters99}.  
Alternatively, upper limits on the stellar M/L can be 
derived from estimating the M/L required to maximise 
the contribution of the stars to the observed rotation curve:
this is called the {\it maximum disc} stellar M/L estimate
\cite{vanalbada85,swaters99}.

\begin{figure*}
\begin{minipage}{175mm}
\begin{center}
  \leavevmode
  \epsfxsize 175mm
  \epsffile{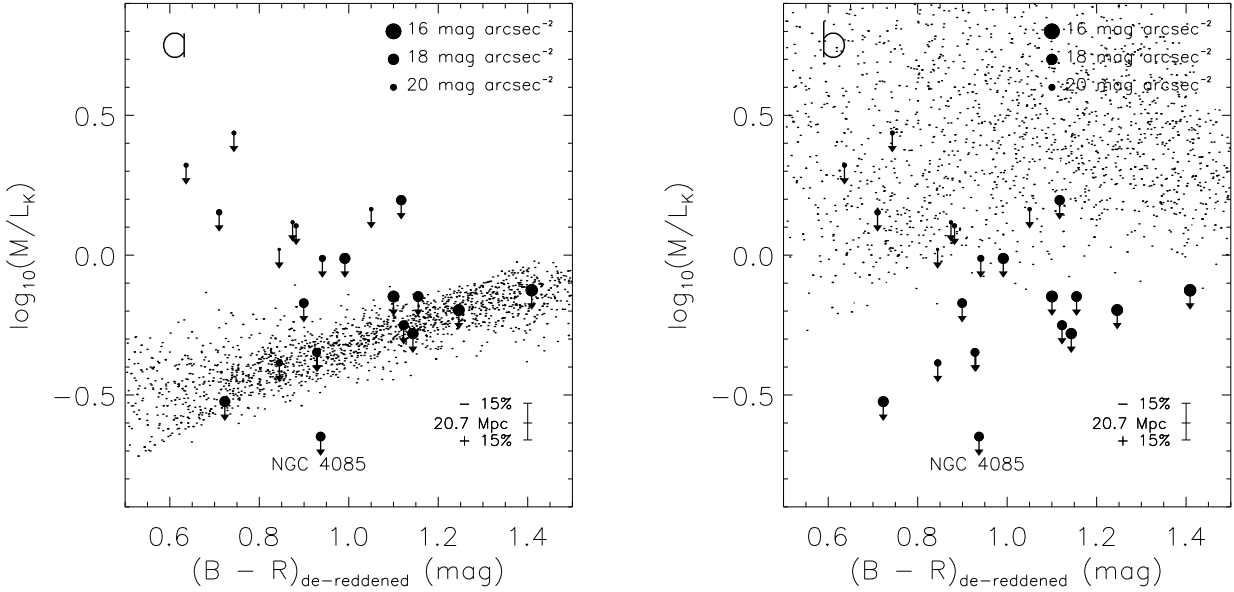}
\end{center}
\caption{ (a) Observed $K$-band {\it stellar} M/L, plotted against
de-reddened $B - R$ colour. The observational points (solid circles)
give upper limits to the stellar M/Ls and were inferred using the
assumption of a maximum disc. The dots show model stellar M/L.  The
data come from $K$ band imaging and \protect\hi rotation curves
(Verheijen 1997; his Chapter 6) of galaxies in the Ursa Major cluster.
The effect on the maximum disc M/Ls of a $\pm 15$ per cent Ursa Major Cluster
distance error is also shown.  NGC\,4085 is highlighted: it has a
poorly resolved rotation curve, which biases the maximum disc estimate
of M/L downwards.  Symbol size is coded by inclination-corrected
$K$-band central surface brightness, as indicated in the legend.  (b)
Total (rather than stellar as in (a)) $K$-band M/L for model galaxies
(dots). The M/L values inferred from the observations are the same as
in (a). }
\label{fig:ml}
\end{minipage}
\end{figure*}

Here, we focus on comparing model stellar M/Ls with observations of
maximum disc stellar M/Ls.  This is facilitated by recent results from
Bell \& de Jong \shortcite{ml}, who argue that stellar M/L is
primarily a function of colour, almost completely independent of SFH
and model uncertainties.  In Fig.\ \ref{fig:ml}(a), we compare the
observed $K$-band maximum disc stellar M/L estimates for spiral
galaxies in the Ursa Major Cluster \cite{verheijen} with stellar M/Ls
from the fiducial model of CLBF, as a function of $B - R$ colour.  The
Ursa Major Cluster is assumed to have the HST Key Project distance of
20.7 Mpc; accordingly, for this 
discussion, we adopt the HST Key Project value of Hubble's
constant, ${\rm H}_0$, of $71$\,km\,s$^{-1}$\,Mpc$^{-1}$
\cite{sakai00}.

An important point is that the stellar M/L of a given 
observed galaxy cannot exceed its maximum disc value.  
Increasing the amount of dark matter in the inner parts of 
a galaxy requires a corresponding decrease in the stellar M/L, 
as shown by the arrows.
The lower envelope of the observed galaxies is the key constraint:
observed galaxies with maximum disc stellar M/Ls substantially larger than
this lower envelope are likely to be dark matter dominated, primarily
LSB, galaxies.  The rotation curve of NGC 4085 suffers from substantial
beam smearing, which artificially lowers the observed M/L.  We therefore 
ignore this galaxy in the following discussion.  
Further discussion of the sources of error in, and the 
implications of, this plot are presented in Bell \& de Jong \shortcite{ml}.

From Fig.\ \ref{fig:ml}(a) it is clear that the model galaxies have 
stellar M/Ls which are consistent with the M/L values inferred 
using the maximum disc assumption (in agreement with CLBF's 
less detailed analysis of spiral discs).  
However, galaxies in the semi-analytic model are in fact poorly 
described by the maximum disc model as they have a significant 
amount of dark matter within the half-light radius of the disc. 
This is clearly shown in Fig.\ \ref{fig:ml}(b), in which we plot the 
M/L for model galaxies using a dynamical estimate of the mass contained 
within the half mass radius, derived from the rotation speed. 
The inclusion of the dark matter component increases the scatter 
in the model M/L values and also shifts the lower envelope 
upwards by around 60 per cent. 
This effect is similar in size to the offset in the zero-point of 
the TF relation.  Possible solutions
to this discrepancy are discussed later in section \ref{disc:tf},
and include a reduced amount of dark matter in the inner parts
of galaxies, and a reduced stellar M/L.

\section{The spin distribution of spiral galaxies} \label{sec:spin}

The distribution of disc scale-sizes, as a function of luminosity, is
a powerful test of the assumed distribution of initial halo angular
momentum, under the assumption of angular momentum conservation 
during disc formation.  The disc sizes were calculated by 
equating the disc specific angular momentum with the mean 
specific angular momentum of the cooled gas, accounting
for the dark matter and gas profiles, disc self-gravity, 
the gravitational influence of the halo and bulge, and 
adiabatic contraction of these components as a response
to disc formation \cite[CLBF]{mmw}. 
The form of the distribution of halo specific angular
momentum has been firmly established by $N$-body simulations
(e.g., Barnes \& Efstathiou 1987; Efstathiou et al.\ 1988; 
Zurek, Quinn \& Salmon 1988). The
logarithmic width of the distribution of the dimensionless spin
parameter, $\lambda$ is $\sigma_{\lambda} \sim 0.5$, essentially
independently of galaxy halo parameters
\cite{barnes87,frenk88,cole96,lemson99,bullock01}.  
At low luminosities, the scale-size
distribution also tests the feedback model. In fact, the strength of
feedback is tuned, in part, to reproduce the observed trend in the
scale-size distribution as a function of luminosity (CLBF).

The distribution of scale-size as a function of luminosity was
tested by both CLBF (their fig.\ 8) and de Jong \& Lacey 
\shortcite[their fig.\ 10]{dejong00} against de Jong \& Lacey's
\shortcite{dejong00} 1000-galaxy $I$-band dataset. 
Both studies found a reasonable match between model predictions and
observations.  The median scale lengths, and the trend in median scale
length with luminosity, are well-reproduced by the model.  However,
the model distribution is somewhat wider than the observed
distribution: in the model of CLBF, $\sigma_{\lambda} \sim 0.5$,
whereas the observations indicate $\sigma_{\lambda} \sim 0.35$.
Because $\sigma_{\lambda}$ is believed to be independent of halo
properties, this is not a trivial discrepancy to rectify. Possible
solutions were discussed by de Jong \& Lacey \shortcite{dejong00}, and
include disc instabilities and the adoption of thresholds for star
formation and feedback (e.g.\ Martin \& Kennicutt 2001).

Since then, there have been three papers which have published determinations
of the width of the scale-size 
distribution: Cross \& Driver \shortcite{cross02} 
with a 45000-galaxy sample from the Two degree Field Galaxy Redshift Survey
\cite{colless01}, and
Kauffmann et al.\ \shortcite{kauf02} and Shen et al.\ \shortcite{shen}
with $\sim$120000-galaxy
samples from the Sloan Digital Sky Survey \cite{york00}.  
Whereas the  
hybrid stellar mass--$z'$-band determination of Kauffmann et al.\ (2003) 
yielded a distribution with a width equivalent 
to $\sigma_{\lambda} \sim 0.45$, 
Cross \& Driver's \shortcite{cross02} 
photographic $b_J$ sample gave a spin distribution 
width of only $\sigma_{\lambda} \sim 0.25$!  Shen et al.'s \shortcite{shen}
distribution showed $\sigma_{\lambda} \sim 0.35-0.45$ in $r$-band,
with $\sim L^*$ galaxies having lower valies of $\sigma_{\lambda}$.
We note that Shen et al.\ present simple model-based interpretation
of this distribution (in the spirit of Mo et al.\ 1998),
which we do not discuss further here.

However, these seemingly discrepant results (with widths of 0.25 in
$b_J$-band, $\sim$0.35 in $r$-band around $L^*$ 
and in $I$-band, and 0.45 in terms of stellar mass) can,
in fact be explained if variations in the stellar populations and the
dust content of spiral galaxies are taken into account.  The
scale-size--luminosity distribution can be thought of as a surface
brightness--luminosity distribution, by noting that $L \propto \Sigma
r^2$, where $L$ is the luminosity, $\Sigma$ is the surface brightness
and $r$ is the scale-size.  Thus, effectively, what is sought is the
width of the stellar surface density--stellar mass distribution,
whereas what is observed is the surface brightness--luminosity
distribution.  It is well-known that there is a correlation between
surface density and SF history, such that at low surface density,
galaxies tend to have rather younger mean stellar ages (and therefore
lower stellar M/Ls, especially so in bluer bands; Bell \& de Jong
2001) than higher surface density galaxies \cite{papii,kauf02}.
Furthermore, higher surface density galaxies will tend to have
somewhat more dust because of the combined effect of higher surface
density and higher metallicity \cite{dejong00,radfir}.  Thus, at high
surface density the surface brightness becomes somewhat lower, and for
low surface density the surface brightness becomes somewhat higher
(both trends are stronger in the bluer bands).  Thus, the width of the
surface brightness--luminosity distribution will be decreased compared
to the surface density--mass distribution, particularly so in the
bluer bands (as is observed, with $\sigma \sim 0.25$ in $b_J$, 0.35 in
$I$, and 0.45 in terms of stellar mass).

Quantitatively, the magnitude of this effect can be crudely
estimated using the results of Bell \& de Jong \shortcite{ml}
and the surface brightness-dependent dust correction of 
de Jong \& Lacey \shortcite{dejong00}.  From the models of 
Bell \& de Jong \shortcite[see their Fig.\ 1]{ml}, 
$\log_{10} M/L$ in the $B$ and $I$ bands is roughly proportional
to $-1/8 \mu_K$ and $-1/16 \mu_K$, respectively, where
$\mu_K$ denotes the surface brightness in the $K$-band. 
Using the $I$-band as an example, the surface brightness is given by:  
$\mu_I \propto -2.5 \log_{10} \Sigma_I = -2.5 \log_{10} \Sigma + 2.5 \log_{10}
M/L_I$, where $\Sigma_I$ is the luminosity surface density in $I$-band, 
and $\Sigma$ is the mass surface density.  
Taking $\mu_K \sim -2.5 \log_{10} \Sigma = \mu$, where
$\mu$ is the stellar surface mass density (this assumption makes only a 
$\ll 5$\% difference to the final answer), $\mu_I \propto  0.85 \mu$ 
(the corresponding result in $B$-band is $\mu_B \propto 0.7 \mu$).
Thus, differences in stellar M/L alone across the surface brightness
distribution cause the distribution to decrease in width by 15\% in $I$-band
and 30\% in $B$-band.  The effect of dust is less trivial to estimate.
de Jong \& Lacey \shortcite{dejong00} dust-correct their data 
to a face-on aspect, making the impact of dust extinction 
modest in this case and therefore relatively easy to deduce. 
de Jong \& Lacey \shortcite{dejong00} estimate a 5\% effect on the $I$-band
half-light surface brightness, when going from an inclination of 60 degrees 
to face-on. It is perhaps not absurd to postulate a further 5\% effect
when going from face-on, extinguished, to face-on, dust-corrected.  Using 
the extinction curve of Gordon et al.\ \shortcite{gordon97} to estimate
the effect in the $B$-band (a factor of 2.75 greater than in the $I$-band), 
and noting that Cross \& Driver \shortcite{cross02} do not extinction 
correct their data
(and, therefore, the correction from 60 degrees to unextinguished is 
perhaps more appropriate), leads to an estimate of around a 30\% 
effect in $B$ band.  Thus, the observed $I$-band and $B$-band widths
of $\sim 0.35$ and $\sim 0.25$ correspond to 
``mass'' widths of $\sim 0.35 \times 1.05 / 0.85 = 0.43$ for $I$-band, 
and a $B$-band ``mass'' width of $\sim 0.25 \times 1.3 / 0.7 = 0.46$. 
Both values are fully consistent with Kauffmann et al.'s \shortcite{kauf02}
``stellar mass'' value of $\sim 0.45$ and the
theoretical estimates of $\sigma_{\lambda} \sim 0.5$.
It should be borne in mind that these are rough estimates of this
effect; more systematic investigation of the width of the
luminosity--surface brightness (or scale-size) distributions
as a function of passband will further test and refine this crude
correction (Bell, McIntosh, Katz \& Weinberg 2003, in preparation).

\section{Discussion} \label{sec:disc}

Many of the properties of spiral galaxies are well-matched by
the models of CLBF.  For example, the gas fractions are quite well-matched,
with only a small excess of gas-poor galaxies, the metallicities are 
matched to within the errors, the model scale sizes reproduce the 
data well, the optical--near-IR colours as a function of gas fraction
are accurate, and the present-day SFRs are
a good match to the observational data.  Nevertheless, 
the keenest insight into the models is gained when exploring 
aspects that poorly match the observations.
Therefore, we focus in this section on 
the discrepancies between the fiducial model and the observations.
Sections \ref{disc:mag} discusses the deficit of the most luminous
model spiral galaxies, \ref{disc:col} discusses the 
`backwards' optical colour--magnitude relation of model spiral galaxies, and 
section \ref{disc:tf} explores the relationship between the 
model stellar M/Ls and the TF relation.

\subsection{The deficit of luminous model spirals} \label{disc:mag}

\begin{figure}
\begin{center}
  \leavevmode
  \epsfxsize 8cm
  \epsffile[18 195 592 618]{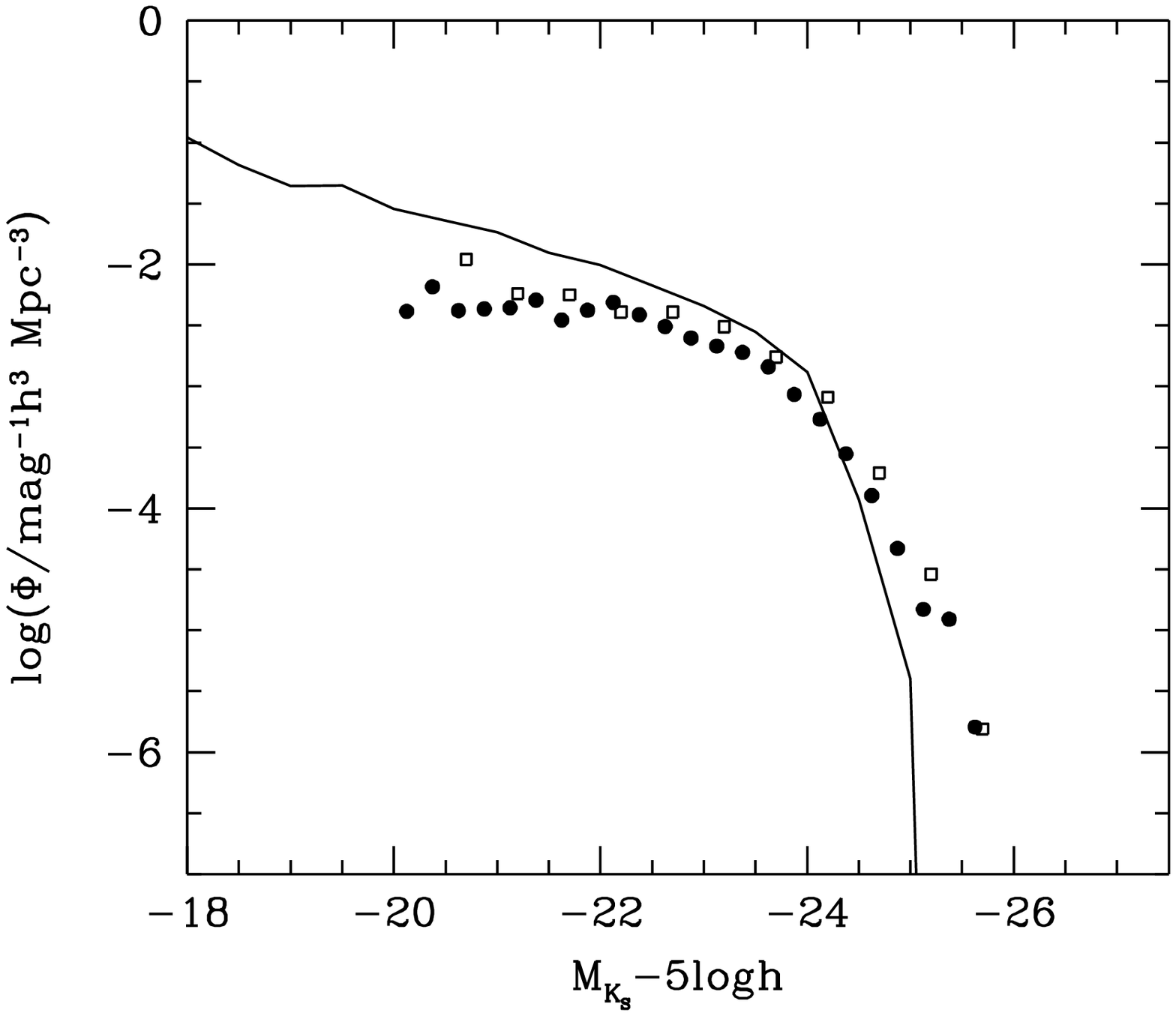}
\end{center}
\caption{ The luminosity function of late-type
galaxies.  Observed $K$-band luminosity functions
for late-type galaxies are shown by the open
\protect\cite{kochanek01} and filled symbols \protect\cite{bell03lf}.
The difference between the observations gives some indication of typing
uncertainty.  The model luminosity function is shown by the 
solid line. }
\label{fig:lf}
\end{figure}

\begin{figure*}
\begin{minipage}{175mm}
\begin{center}
\hbox{%
  \epsfxsize 67mm
  \epsffile[18 195 592 618]{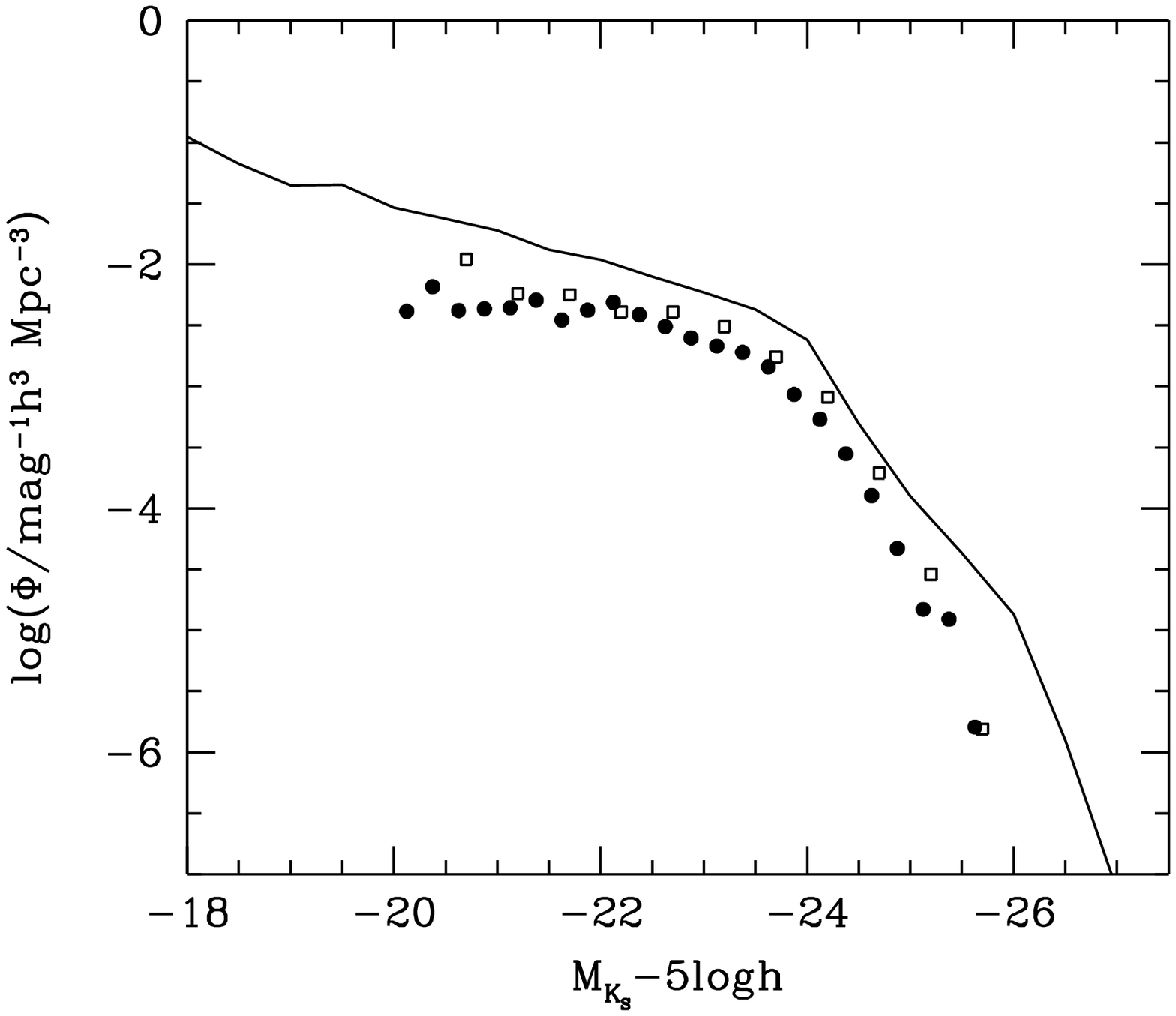}
  \epsfxsize 54mm
  \epsffile{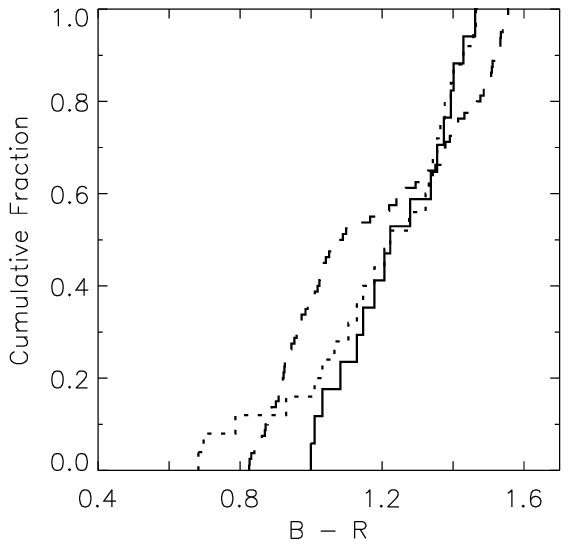}
  \epsfxsize 54mm
  \epsffile{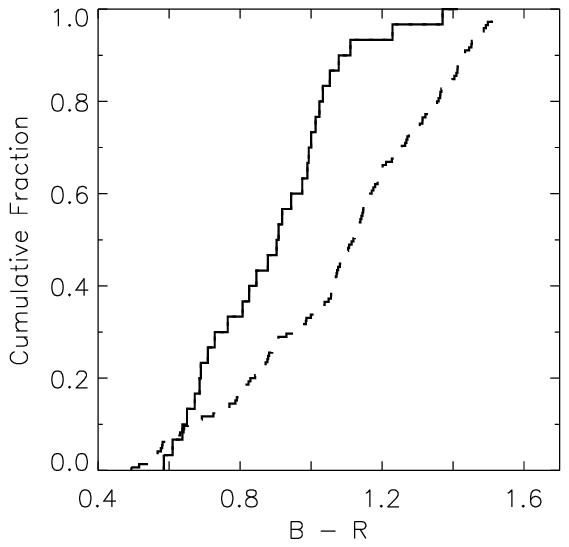}}
\end{center}
\caption{The effects of changing the $B/D$ cutoff to 
$B/D < 100$.  The left panel shows the luminosity function
of late-type galaxies, following Fig.\ \ref{fig:lf}.
The middle and right panels show the cumulative distribution
of $B-R$ colours for all observed galaxies (dotted line), 
observed galaxies with matching model galaxies (solid line), 
and model galaxies (dashed line).  The middle panel 
shows the colour distribution of the most luminous galaxies
only $-26 \le M_K - 5\log_{10} h < -23.5$ (the observations
and models are different at only 95 per cent confidence) and 
the right panel shows the colour distribution of the
faintest galaxies  $-20.5 \le M_K - 5\log_{10} h < -17$.
}
\label{fig:bd100}
\end{minipage}
\end{figure*}

\begin{figure*}
\begin{minipage}{175mm}
\begin{center}
\hbox{%
  \epsfxsize 67mm
  \epsffile[18 195 592 618]{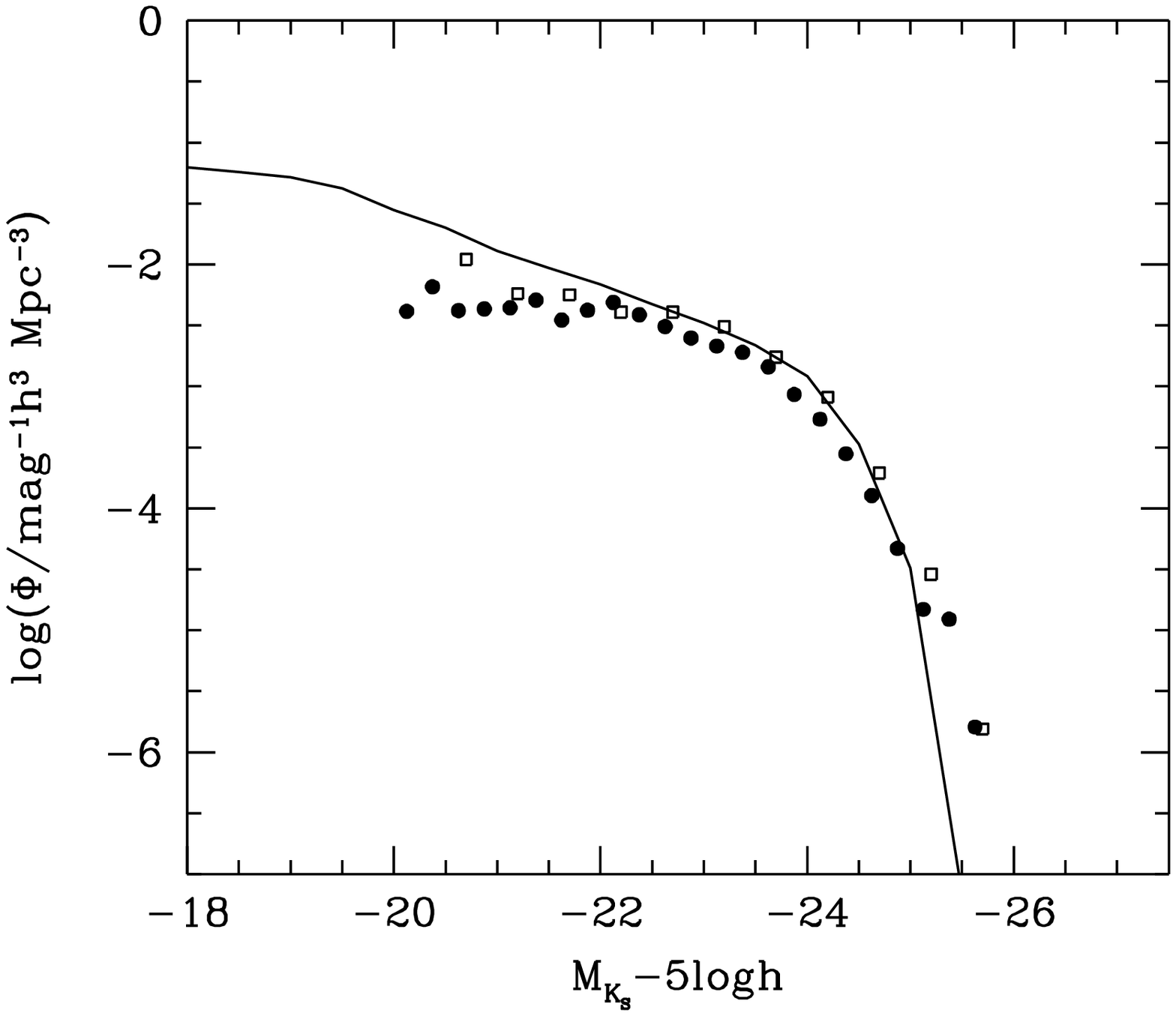}
  \epsfxsize 54mm
  \epsffile{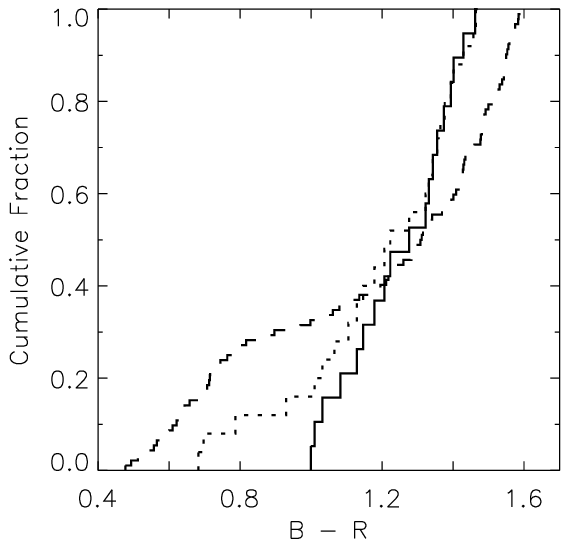}
  \epsfxsize 54mm
  \epsffile{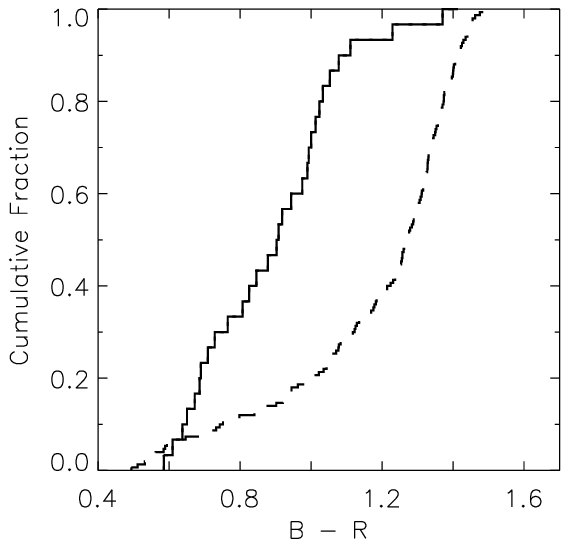}}
\end{center}
\caption{The predictions from a variant of the CLBF model, which has a higher
baryon fraction and a feedback model that includes ``superwinds'' (see
text for further details).  The left panel shows the luminosity function
of late-type galaxies, following Fig.\ \ref{fig:lf}.
The middle and right panels show the cumulative distribution
of $B-R$ colours for all observed galaxies (dotted line), 
observed galaxies with matching model galaxies (solid line), 
and model galaxies (dashed line).  The middle panel 
shows the colour distribution of the most luminous galaxies
only $-26 \le M_K - 5\log_{10} h < -23.5$ (the observations
and models are different at only 95 per cent confidence)and 
the right panel shows the colour distribution of the
faintest galaxies  $-20.5 \le M_K - 5\log_{10} h < -17$.
}
\label{fig:super}
\end{minipage}
\end{figure*}

Both the datasets of de Jong \& Lacey \shortcite{dejong00} and Bell \&
de Jong \shortcite{papiii} support the view  that the models of
CLBF under-produce luminous, large, spiral galaxies.  We explore this
deficit in more detail in Fig.\ \ref{fig:lf}.  Two observational 
determinations of the $K$-band luminosity function of late-type 
galaxies are shown by open symbols \cite{kochanek01} and 
filled symbols \cite{bell03lf}.  The two studies morphologically
classify their galaxies using totally different methodologies.
Kochanek et al.\ \shortcite{kochanek01} classify 
their galaxy samples visually, whereas
Bell et al.\ \shortcite{bell03lf} use the 
Sloan Digital Sky Survey $r$-band concentration
parameter to classify the galaxies, following 
Strateva et al.\ \shortcite{strat}.  The model $K$-band luminosity function 
for galaxies with $B/D \le 1$ in $K$-band is shown by the solid
blue line.  The faint end slope of the observations and 
models are different; at least part of this discrepancy 
is because of incompleteness in the Two Micron All Sky Survey
with respect to low surface brightness galaxies \cite{bell03lf}, 
although some of this discrepancy may be real.
Significantly, the luminosity function of 
galaxies with $K$-band $B/D \le 1$ cuts off sharply
above $M_K - 5\log_{10}h \sim 24$, in clear contrast
to the observational data.  Because the
parameters of the CLBF model are chosen to reproduce the {\it total} observed
local luminosity function closely, it is challenging to construct a
model which produces more luminous spiral galaxies without violating
the constraints of the overall luminosity function.  Because of this,
there are two possible solutions to these problems: 
first, that many of the most luminous model
galaxies are being incorrectly converted into ellipticals; and second,
that the prescription for the accretion of gas by spirals may require
modification.

The first possibility 
is that the underproduction of bright spirals is linked to
deficiencies in the parameterisation of the physics of galaxy mergers.
The mass ratio at which (major) mergers transform spirals into
ellipticals is tuned to reproduce the morphological mix of galaxies
brighter than $L_*$.  Nevertheless, there are a number of
bulge-dominated galaxies in haloes less massive than 10$^{14} {\rm
M_{\sun}}$ with $B/D > 1$ and with magnitudes up to $M_K - 5
\log_{10}h = -25.1$.  If mergers were, for some reason, less effective
in transforming spirals into ellipticals, it may be possible to
produce a larger population of red, massive spirals.  This could be
achieved by altering the prescription for redistributing the disc
stars after a major merger; for example, a good fraction of the
pre-existing stellar population could be transferred to a thick disc
instead of being assigned to a spheroid component (see, e.g.,
Dalcanton \& Bernstein 2002).  Alternatively, the dynamical friction
calculation of orbital decay may underestimate the merger timescale
due to the neglect of mass stripping from the satellite \cite{benson02}.  
To explore the possibility that
galaxies could be mislabeled in the model, we extend the comparison
with observation in Fig.\ \ref{fig:bd100} to include model galaxies
with $B/D < 100$.  Using this extreme cut-off, 
the model produces as many luminous galaxies as required (to within
the typing uncertainties).  Furthermore, 
these additional galaxies are very red, 
bettering the match between model and observed luminous 
spiral galaxies.  Such an extreme shift in the boundary of the $B/D$ ratio
demarcating spirals and ellipticals is a highly unattractive option,
however, as this would grossly violate the constraints on the
morphological mix of bright galaxies, and substantially under-produce
the observed luminosity function of early-type galaxies.

The other possibility is that the prescription for accretion of cool
gas by spiral galaxies may require modification.  This could be
modelled by changing the gas cooling prescription. Indeed, when CLBF
changed the density profile of cooling gas, the bright end of the
luminosity function was substantially affected (their fig.\ 6, panel
c).  The $\beta$-profile adopted by CLBF was chosen as it is in good
agreement with the results of numerical simulations \cite{eke98}.
Furthermore, comparisons of gas cooling in semi-analytic and
gas-dynamical simulations have, rather encouragingly, given, similar
results overall \cite[although see Katz et al.\ 2002 for a differing
viewpoint]{benson01a,helly02b}.  An interesting and potentially
fruitful approach (which is beyond the scope of this paper) would be
to search for a physical motivation for a modification of the gas
cooling which allows more gas inflow at earlier times (to produce
brighter, discy galaxies) and less infall now (to produce redder
bright galaxies) to address both the under-production of large, bright
spiral galaxies and the rather blue colours of bright model spiral
galaxies.

Another way to vary the amount of gas accreted by spiral galaxies is
to modify the feedback prescription.  We examine a variant of the CLBF
fiducial model in Fig.\ \ref{fig:super}.  In this new model, the
baryon density is twice the value used by CLBF, in line with recent
constraints on primordial nucleosynthesis inferred from quasar
absorption line studies \cite{omeara01,spergel03}.  
A further key change is that
some fraction of the gas reheated by feedback is permanently ejected
in a ``superwind'' \cite{benson03}.  The
removal of baryons by the superwind has the consequence that fuel is
removed from the more massive galaxies that form later on in the
hierarchy.  The superwind model does somewhat better than the fiducial
model of CLBF at reproducing large scale length, bright
spirals.  This strong feedback delays 
gas cooling to somewhat later epochs, yielding larger
galaxy sizes.  The feedback does not delay this gas cooling
too long, however, and indeed is efficient enough to substantially suppress
very late gas cooling, yielding relatively red colours for these
large scale-length galaxies.
Typically, models with a ``high'' baryon fraction
($\Omega_{b}h^{2}=0.02$ as opposed to the value of
$\Omega_{b}h^{2}=0.01$ adopted in the fiducial model of CLBF) require
a significant fraction of brown dwarfs to be invoked in order to match
the luminosity function at $L_{*}$. This is not the case in the
superwind model, which predicts M/Ls in line with those presented in
Section \ref{sec:ml} (because baryons that would have been locked
up into brown dwarfs in a high baryon fraction model without superwinds
are incorporated in these energetic winds and lost forever in the superwind
model).  A significant challenge to this model, however, 
is the stringent energetic requirement: 
the amount of energy required to eject the gas exceeds
that available in stellar winds and supernovae, necessitating
a significant contribution from active galactic nuclei \cite{benson03}. 

\subsection{The colour--magnitude relation of spiral galaxies} \label{disc:col}

Arguably the most significant discrepancy between the models and the 
data is the `backwards' colour--magnitude relation.  The data
clearly indicate that bright spiral galaxies are redder in the optical
than fainter spiral galaxies.  In contrast, the models predict that 
fainter spiral galaxies should be slightly redder than 
bright spiral galaxies (Figs.\ \ref{fig:colcolmag} and \ref{fig:colcum}).  
This result is robust to {\it modest} changes in the $B/D$ limits we 
adopt, up to $B/D < 10$.  A more sophisticated dust
model is unlikely to help in this respect, as we saw earlier that
varying the dust prescription produced only a very minor effect
(much smaller than the discrepancy discussed here), and in any 
case would not address the red colours of the faint model disc galaxies.
That the models should predict
that the colours of luminous spiral galaxies would be bluer
than their fainter counterparts is relatively easy to understand.
Hierarchical galaxy formation models predict that small halos 
are assembled before larger haloes; therefore, it is natural
to find that more massive galaxies in larger haloes should
have bluer colours, indicative of younger stellar populations, than
those of less massive galaxies in smaller haloes.  Thus, the 
challenge is to find a mechanism which can override
this natural sequence of formation by ensuring that
the stellar populations ending up in more massive 
spirals today can form earlier than those ending up in
less massive spiral galaxies.
We discuss three possible remedies for this discrepancy: 
varying the definition of the model spiral galaxies, varying
the prescription for turning gas into stars, and varying the 
amount of gas in the galaxies in the first place.

One of the options discussed in section \ref{disc:mag}
for increasing the numbers of luminous spiral galaxies was
making morphological transformation from a disc-dominated to 
a spheroidal galaxy more difficult.
This increases the number of red galaxies in the most luminous bin
(the right hand panel of Fig.\ \ref{fig:bd100}), in better
accord with the observations.
This comes, however, 
at the cost of violating constraints set by the morphological
mix of $L_*$ galaxies, making this option relatively unattractive.
Furthermore, this does not affect the faint model galaxy colours,
which remain too red.

Another possible way to both redden the most luminous model
galaxies and make the faintest model galaxies bluer is to modify
the SF law.  Indeed, Bell \& Bower \shortcite{papiii} argued
that a SF law of the type used here may imprint a weak `backwards' optical
colour--magnitude relation (brighter galaxies blue and fainter
galaxies red).  However, tests that we have carried out have shown 
that the use of (for example) a density-dependent
star formation law does not significantly affect the colours
of spiral galaxies in this model.  Furthermore, van den Bosch
\shortcite{vdb02} adopts a density-dependent Schmidt SF law in 
his semi-analytic model, and finds, if anything, a stronger 
`backwards' colour--magnitude relation for spiral galaxies than the
model explored in this paper.

Perhaps the most promising possibility for altering the colours of the
brightest spiral galaxies is to modify the gas accretion prescriptions
(as discussed above; this is also the favoured solution of van den
Bosch 2002).  Varying the amount of gas that cools onto luminous model
galaxies can substantially change their recent SFH (as well as their
luminosities; CLBF). However, a detailed exploration of this
possibility is beyond the scope of this paper.  Varying the amount of
gas that can cool today through modifications to the feedback
prescription (the superwinds model) does clearly produce a number of
optically red, large scale length spiral galaxies, in much better
agreement with the observations.  However, as is expected for a model
with very strong feedback at all galaxy masses, the
metallicity--magnitude relation is somewhat too shallow (the
metallicity, as measured by $R - K$ colour, is too low for the
luminous spiral galaxies; not shown).  Another significant
challenge is that the strong feedback makes faint disc-dominated
galaxies even redder than the fiducial model, exacerbating 
the already poor match between the red model galaxies and 
the observed blue galaxies.
Further work is required to explore the origin
of the spiral galaxy colour--magnitude relation in more detail.

\subsection{Mass-to-light ratios and dark matter halo profiles} \label{disc:tf}

In section \ref{sec:ml}, we discussed the {\it stellar} M/Ls of model
galaxies from CLBF, finding that they are as large as can be plausibly
accommodated by the data.  Thus, in order for the model galaxies to
satisfy observed maximum disc constraints they must have little dark
matter in their inner parts.  However, in the fiducial model,
typically 60 per cent of the mass in the inner half-light radius of
the galaxy is non-baryonic dark matter (Fig.\ \ref{fig:ml}(b)).  This
is exactly the offset that the models have from the observed TF
relation: the rotation speed (which is proportional to the 
square root of the enclosed
mass) is too high at a given luminosity.  This is the origin of the
discrepancy between the good match to the luminosity function and the
poor match to the zero-point of the TF relation. 

This discrepancy could be alleviated by lowering the stellar M/Ls to
become substantially sub-maximal, in concert with reducing the overall
mass of baryons that end up in discs.  This is achieved by CLBF in
their $\Omega_b h^2 = 0.005$ model (which was considered by CLBF 
only to illustrate the effects of parameter changes on the results), 
which, because it is normalised to
reproduce the luminosity function of galaxies, needs to produce the
same amount of light with a smaller baryonic mass, leading to a lower
stellar M/L.  It is worth noting that this baryon density is four times
lower than the current best estimate for the cosmic baryon density
from both Big Bang nucleosynthesis \cite{omeara01} 
and the cosmic microwave background \cite{spergel03}, 
and is strongly ruled out --- we discuss this case to illustrate
lower stellar M/Ls only.  This lower baryonic mass in the disc leads to less
adiabatic contraction of the halo. For both reasons, disc circular
velocities are lowered, leading to a smaller offset from the TF
relation.  The level to which all discs, even HSB spiral galaxies,
could be sub-maximal is a matter of intense debate (see e.g.\ Courteau
\& Rix 1999; Weiner et al.\ 2001 and references therein).  The
tightness of the lower envelope of the observed stellar M/L--colour correlation
in Fig.\ \ref{fig:ml} suggests that at least the highest
surface brightness galaxies have to be a roughly constant fraction of
maximal (of course, maximal discs automatically adhere to that
constraint).

Alternatively, the discrepancy could be alleviated by reducing the
amount of dark matter in the inner parts of the galaxy, by for example
changing the dark matter halo profile to have a flatter core.  This
has been suggested a number of times in the past, to better explain
e.g.\ the rotation curves of LSB galaxies \cite{deblok01}, or lensing
constraints on kpc scales in elliptical galaxies \cite{keeton01}.  An
{\it ad hoc} modification of this kind is relatively unattractive, at
least from a theoretical point of view, given the strength of $N$-body
evidence indicating that cold dark matter haloes should be at least as
concentrated as an NFW profile (NFW, Ghigna et al. 2000).  Alternatively,
adopting an alternative type of dark matter, such 
as warm dark matter (which leads to less concentrated haloes)
may also solve this problem \cite{eke01}.

\section{Conclusions} \label{sec:conc}

We have compared the properties of local spiral galaxies with model
spiral galaxies from the simulations of CLBF while taking into account
observational selection and striving to construct model and observational
quantities in very similar ways.  In this fashion, we have undertaken
a detailed exploration of baryonic processes such as gas cooling and
the SF law.

Model successes include the following:
\begin{itemize}
\item Model galaxy magnitudes, scale-sizes and gas fractions are in 
reasonable accord with the data (although there
is a small excess of gas-poor model galaxies).  Particular successes include
a tight gas fraction--surface brightness correlation, and the variation 
in mean scale-size with luminosity.  
\item We demonstrate that the seemingly narrow ($\sigma \sim 0.3$) 
spiral galaxy scale-size
distribution derived from optical data is consistent with
the the theoretically-favoured
$\sigma \sim 0.5$ halo spin distribution, when one takes into account
variations in stellar populations and dust content.
\item The model metallicity--magnitude relation is consistent with 
the data, to within the systematic uncertainties.  
\item The model optical--near-IR colours match trends in 
the observed dataset well as a function of
gas fraction, and to a certain extent scale-size and surface brightness.
\item The present-day SFRs of the model galaxies are in reasonable 
agreement with the data.  We find that it is practically impossible to
differentiate between star formation laws that depend on density or on
dynamical time based on global present-day SFRs alone.
\item Fiducial model stellar M/Ls are consistent with maximum disc
constraints of local spiral galaxies. Substantially higher model 
stellar M/Ls are difficult to obtain using current stellar population
synthesis models.
\end{itemize}

There are some interesting model shortcomings, which offer 
considerable insight into the models:
\begin{itemize}
\item There is a deficit of luminous model spiral galaxies.  
Furthermore, the model
has a `backwards' colour--magnitude relation.  An attractive possibility
for resolving these discrepancies is the adoption of a gas cooling
model that would lead to enhanced (suppressed) infall at earlier times 
for larger (smaller) galaxies.  First promising steps
have been taken by adopting a superwind model in which some fraction of the
gas reheated by feedback is permanently ejected, thus suppressing
cooling rates at low redshifts for bright galaxies.  However,
the faint galaxies become even redder, owing again to this
strong feedback.  Another possible solution might be a more
accurate treatment of galaxy mergers which might suppress the number
of major mergers that lead to the formation of elliptical galaxies. 
\item The model produces stellar mass-to-light ratios which are similar to the 
values derived for many galaxies under the ``maximum disc
hypothesis.'' Yet, there is a substantial dark matter component within
the half-light radius of the model galaxies.  This produces model
discs which rotate too fast at a given luminosity, giving rise to the
well-known offset between the observed and model TF relations.
Possible resolutions of this discrepancy include lowering the model
stellar M/Ls to substantially sub-maximal values, in concert with a
reduction in the mass of baryons that ends up in galaxy discs, or
modifying the dark matter halo profiles so that they have less dark
matter in the inner parts of the galaxy. 
\end{itemize}

\section*{Acknowledgements}

We acknowledge useful conversations with Vincent Eke and Roelof de
Jong.  The referee, Frank van den Bosch, is thanked for helpful
and constructive comments which improved the quality of this paper.
This work was supported by the European Research Training
Network {\it Spectroscopic and Imaging Surveys for Cosmology}.  E.\
F.\ B.\ was also supported by NASA grant NAG5-8426 and NSF grant
AST-9900789.  C.\ M.\ B.\ acknowledges a Royal Society University Research
Fellowship.  This project made use of STARLINK facilities in Durham
and was supported in part by the PPARC rolling grant in Extragalactic
Astronomy and Cosmology at Durham.

\end{document}